\documentclass[12pt,preprint]{aastex}
\def \etal{{\it et~al.}}

\def \ksM{km~s$^{-1}$~Mpc$^{-1}$}

\def\mV{V$_{\rm 606}$}
\def\mI{I$_{\rm 814}$}
\def\mJ{J$_{\rm 110}$}
\def\mH{H$_{\rm 160}$}

\def\H0{H$_{0}$}
\def\q0{q$_{0}$}
\def\Msun{\ifmmode M_{\odot} \else $M_{\odot}$\fi}

\received{2002 October 3}
\shortauthors{Storrie-Lombardi, Weymann, and Thompson }
\shorttitle{HDF High Redshift Candidates and Small Galaxies}
 
\begin{document}

\title{High Redshift Candidates and the Nature of Small Galaxies in the Hubble Deep Field} 
 
\author{Lisa J. Storrie-Lombardi}
\affil{SIRTF Science Center, Caltech, MS 220-6, 
Pasadena, CA 91125; ~lisa@ipac.caltech.edu}

\author{Ray J. Weymann}
\affil{Observatories of the Carnegie Institution of Washington,
813 Santa Barbara Street, Pasadena, CA 91101; rjw@ociw.edu}
\and

\author{Rodger I.\ Thompson}
\affil{Steward Observatory, The University of Arizona, Tucson, AZ 85721;
rthompson@as.arizona.edu}

\footnotetext[1]{Based on observations obtained with the
Near-Infrared Camera and
Multi-Object Spectrometer and the Wide Field and Planetary Camera 2
on the NASA/ESA Hubble Space Telescope
which is operated by AURA Inc., under contract with NASA.}
 
\begin{abstract}
We present results on two related topics: 1) A discussion of high
redshift candidates ($z > 4.5$) and 
2) A study of very small galaxies
at intermediate redshifts, both sets being detected in the region of the northern
Hubble Deep Field covered by the deep NICMOS observations
at 1.6 and 1.1 microns. The high redshift candidates are just those with 
redshift $z > 4.5$ as given in the recent catalog of Thompson, 
Weymann and Storrie-Lombardi,
while the ``small galaxy'' sample is defined to be those
objects  with isophotal area $\le 0.2 ^{\Box ''}$ and  with photometric
redshifts $1 \le z \le 4.5$.  

Of the 19 possible high redshift candidates listed in the Thompson \etal\ catalog,
11 have (nominal) photometric redshifts less than 5.0. Of these, however, only
4 are ``robust'' in the sense of yielding high redshifts when the fluxes are
randomly perturbed with errors comparable to the estimated measuring error in
each wave band. For the 8 other objects with nominal photometric redshifts
greater than 5.0, one (WFPC2 4--473) has a published spectroscopic redshift. Of
the remaining 7, 4 are robust in the sense indicated above. Two of these form
a close pair (NIC 586 and NIC 107). The redshift of the
object having formally the highest redshift, at 6.56
(NIC118 = WFPC2 4--601), is problematic, since F606W and F814W flux are clearly present,
and the nature of this object poses a dilemma.

Previous work by Colley \etal\ has suggested that compact
sources in the WFPC2 HDF images are subgalactic components at
redshifts $z > 0.5$ since they are correlated on scales less than $1"$,
corresponding to physical scales of less than 8 kpc
(\H0 = 65 \ksM, \q0 = 0.125). We confirm these correlations in the
WFPC2 data. However, we do not detect the correlation of close 
pairs of galaxies on small scales in 
the $\sim 0.65^{\Box '}$ region of the HDF that we surveyed with NICMOS.
The smaller area surveyed and lower resolution will make any real 
correlation more difficult to measure in these data. 
We have examined averaged images of these faint
({\mV} $\sim$ 27--29), compact objects to search for
extended, surrounding flux from older, fainter populations of stars.
We find no evidence from the averaged images that isolated, compact objects in 
the Hubble Deep Field are embedded in fainter, more extended galaxies. 
For three different assumptions about possible star formation histories in these
objects we set limits
on the total amount of stars which could have been formed in an annulus
corresponding to radii between $\sim$ 6 to 10 kpc, which is typically a few
times $10^8\Msun$. We suggest that some of these objects may be protogalactic
fragments.

\end{abstract}


%%%% SECTION 1
\section{Introduction}

The redshifts, number counts, sizes, morphologies, and colors of galaxies
are used to determine their evolutionary history and constrain
cosmological models. It is therefore crucial to understand what
we are counting and measuring as we peer deeper into the
history of the Universe where objects appear fainter, smaller,
and commonly measured features in rest frame optical wavebands move
into the near-infrared and beyond. The optical images
of the Hubble Deep Field (HDF; Williams \etal\ 1996) showed
us an exquisitely detailed view of some high redshift galaxies,
though the view is inherently distorted (Ferguson 1998)
due to the effects of cosmological surface brightness
dimming, the effects of dust on galaxy colors, and the fact
that we view higher redshift galaxies at progressively bluer
rest wavelengths.

NICMOS observations of
the Hubble Deep Field (Thompson \etal\ 1998) provide us with two additional
redder wavebands to help disentangle some of these effects.
In Thompson, Weymann, \& Storrie-Lombardi (2001; hereafter TWS) we presented 
a comprehensive table of objects
detected in these six wavebands, together with estimates for their redshifts,
star-formation rates and internal extinction, and estimated the
global star formation rate as a function of redshift based on these results.
We also commented briefly on some objects of special interest.

In the present paper, we discuss in more detail two different, but related
subsets of objects relevant to the early formation of
galaxies: 1) A set of 19 objects which 
are candidates for being high redshift\footnote{Hereafter we use the phrase ``high redshift'' to refer to values
$ > \, 4.5$ and the phrase ``intermediate redshift'' to refer to values $ 1 \le z \le 4.5$.}
galaxies according to Table 1 of TWS, regardless of their morphological
properties. 2) A set of objects
with estimated redshifts at intermediate values, and which are characterized by
being very small, some of which may be protogalactic
objects which will eventually be assembled into normal galaxies.\footnote{
For reasons discussed below however, there is not quite an 
exact one--to--one correspondence between these two subsets  and the subsets of objects in 
TWS meeting similar criteria.}
%RJWAUG9_E1

The motivation for examining the high redshift candidates is obvious: We would like to
understand the epoch and manner in which the earliest episodes of star formation took place.

The motivation for our study of the set of intermediate redshift, very small objects stems
from a paper by (O'Connell \& Marcum 1997). For these objects, 
the combination of the brighter, clumpier
star--forming regions moving into the optical filters, and
surface brightness dimming  will make these intermediate redshift, 
compact, UV-bright objects more prominent
than lower surface brightness objects at the same redshift.
(Colley \etal\ 1996)  discussed
how this is evident in the Hubble Deep Field by measuring
the two-point correlation function of galaxies detected in the
WFPC2 fields. They found a positive signal in the correlation function for
scales $\le 1$ arcseconds for small objects.
At cosmological distances
this corresponds to subgalactic scales, {\it e.g., } 1 arcsecond $\approx$
8 kpc for redshifts z $>$ 1 (H$_{0}$ = 65 km~s$^{-1}$~Mpc$^{-1}$, q$_{0}$ = 0.125).
This led them to suggest that many
of the `galaxies' detected by source counting algorithms are
probably sub-galactic components. 
Alternatively, some of these very compact objects could well be
small protogalaxies undergoing their first episodes of star formation
even though they are at somewhat lower redshifts than the high redshift
candidates mentioned above. In this case, some models of
galaxy formation suggest that one might {\it not} expect a strong
signal in the two-point correlation on such small scales 
(Rauch, Haehnelt \& Steinmetz 1997 [see pages 603,622]; Steinmetz 1998).
Understanding this issue is obviously relevant in how we interpret
number counts. The two NICMOS bands enhance
our ability to  detect fainter, older stellar populations
that might be connected with brighter star--forming regions
seen at bluer wavelengths, as well as galaxies
whose radiation is attenuated by dust at optical wavelengths.

The paper is organized as follows.
In \S~\ref{sect_hizcand} we examine in more detail the list of high redshift
galaxies listed in Table 1 of TWS and discuss some of the ambiguities that
arise in the interpretation of some of these objects.
In \S~\ref{sect_selection} we discuss how the sample of small, intermediate
redshift galaxies was selected
and the redshift estimates for this sample.
In \S~\ref{sect_average} we discuss averaging the images of fainter galaxies
in this sample in order to 
place limits on the extended flux. 
In \S~\ref{sect_exp_flux} we discuss the expected extended flux 
from compact galaxies at intermediate redshifts, based upon moving observed
lower redshift star--forming galaxies to higher representative redshifts.
In \S~\ref{sect_corr} we  
examine the two-point autocorrelation 
function of the optically selected galaxies with high redshift colors
and the small galaxies selected in the NICMOS field.
In \S~\ref{sect_discuss} we summarize and discuss our main conclusions.

%%%% SECTION 2
\section{High Redshift Candidates}
\label{sect_hizcand}

Examination of Table 1 of TWS reveals that there are 19 objects whose 
photometric redshifts exceed 4.5  
(Note that because of the lookup table in TWS all the photometric redshifts
are quantized in steps of 0.08) In Table~\ref{t_hiz_tab}
we reproduce selected columns of those rows from Table 1 of TWS containing these
19 objects, but now sorted according to the photometric redshifts
published in TWS. 

\placetable{t_hiz_tab}

Column 1 contains the NICMOS ID number, and, when identification
with an object in the Williams \etal\ (1996)  catalog has been made according to the
precepts given in TWS, column 2 contains the WFPC2 identification. Columns 3,4 contain
the total and 0.6 arcsec aperture F160W magnitudes. Columns 5 and 6 are the J2000
coordinates (12h 36m is to be added to the RA; +62 degrees is to be added to the Dec.).
Column 7 gives the photometric redshift published in TWS.
The remaining columns are explained below.
In Figure \ref{f_chisq_panel}
we show, for these 19 objects,  the run of $\chi^2$ vs. redshift for 
the particular extinction and population template selected
by our photometric redshift code. 
\placefigure{f_chisq_panel}
By definition,
the lowest value occurs at redshifts above 4.5. However, in some cases it is apparent
that there are other minima almost as deep at lower redshifts. The absolute value of the
minimum $\chi^2$ fit is not a good indicator of the robustness of the high redshift nature
of these candidates, since (cf TWS equation (1)), very faint objects will tend to have
smaller values of $\chi^2$ than brighter ones for comparable percentage errors in the fitted
and observed fluxes. The relative depth of the primary minimum compared to other minima
give some indication of the robustness of the photometric redshift. However, as in TWS, a 
somewhat better
indication of the robustness of the photometric redshifts can be
obtained by randomly perturbing the fluxes according to the estimates for the distribution
of the photometric errors in each of the 6 wave bands of interest---F300W, F450W, F606W, 
F814W, F110W and F160W, though none of these objects
have any measurable flux in the F450W or F300W bands. However, in contrast to the ``typical'' galaxy in
TWS, most of these 19 objects are extremely faint, and the photometric errors relatively large.
Consequently, the perturbed fluxes yield photometric redshifts which are far less robust than
brighter objects. The distribution of these perturbed photometric redshifts is not remotely Gaussian, or
even symmetrical about the median value, but typically will have a very low redshift
( 0 to $\sim$1) tail in the distribution of the perturbed photometric redshifts, with a discontinuous
jump to high values (typically to  z $ > 4.0$). For each of the 19 objects in the table, 100
randomly perturbed fluxes were generated and run through exactly the same photometric redshift
estimation algorithm described in TWS. For a given object, when these redshifts are ordered, their
ranks can thus be crudely regarded as  confidence estimates for the redshift. In columns 9 and 10
we give the redshifts at which $\sim 5\%$ of the perturbed set have a lower value, and
for which  $\sim 10\%$ of the perturbed set have a higher value.
In view of the comment above about
the discontinuous nature of these redshift distributions, another indicator of robustness is
the rank at which the redshift is above 4.0, and this rank is given in column 11. In the discussion
below we consider that if this rank is 5 or less (ie, less than 5\% of the perturbed redshifts
fell below 4.0) then the classification of the object as a high redshift galaxy is ``robust''.
In column 8 the symbol R is used for the robust objects and NR for the non-robust objects.
We now give a very brief description of the morphology and run of the $\chi^2$ fit with redshift
for each of the robust objects, and, because of its peculiar nature the object with the highest
nominal photometric redshift in the TWS list, NIC118.0.

\noindent {\bf NIC131.0:} \\
There is a well-defined minimum at the published photometric redshift; 
all the other secondary minima are much poorer fits. It is a very faint
smudge on the F606W image, distinctly brighter and resolved on the F814W image with
an indication of a brighter nucleus. Both the F110W and F160W images are slightly
resolved and readily visible, though the lower resolution of the NICMOS images
prevents detailed comparison between the F110W and F160W images and the WFPC2 F606W and
F814W images, a comment that applies to all subsequent objects and will not be repeated.

\noindent {\bf NIC274.0:} \\
While there is a well-defined minimum at z = 4.64, a secondary minimum at
a redshift of 0.80 is almost as deep, and is fit by a hot unreddened population,
whereas the fit at 4.64 requires some reddening of a hot population.
Thus, while this is formally robust against flux perturbations, its reality
as a high redshift object is less convincing than for NIC131.0. The F606W image
appears compact, while the F814W image appears to have a small extension.
The F110W image is bright with no obvious structure, while the F160W image shows
a small nearby companion.

\noindent {\bf NIC267.0:} \\
A well-defined minimum occurs at z = 4.80, and there is no other reasonable
fit. All four images are compact and show
no structure.

\noindent {\bf NIC150.0:} \\
As in NIC267, the $\chi^2$ plot against redshift shows only a deep minimum
at a redshift of 4.80 and no other reasonable fit. There is no WFPC
identification and no object is visible on the F606W image. However there are
two small slightly amorphous objects clearly visible on the F814W image, one of which agrees
well with the nominal F160W and F110W coordinates. However both the F110W and F160W images show
the same two small objects with possibly a more diffuse fainter chain-like 
structure underlying them.

\noindent {\bf NIC277.212:} \\
The primary minimum at a redshift of 5.04 is deep, but rather broad. There is also
a well defined secondary minimum at a redshift of 0.8, though not nearly as good
a fit as the high redshift minimum. The F160W image does not separate this object from
a nearby larger and brighter galaxy, but the F110W image shows it as an apparently
well-separated object. The WFPC2 images are curious, in that the F814W image is quite
bright, but there is no convincing evidence at all of the object in the F606W 
image, which, given the brightness in the F814W image
and the nominal redshift is surprising.
Thus, the nature and redshift of this object is somewhat suspect.

\noindent {\bf NIC184.0 = WFPC2 4--473.0:} \\ 
This object has a spectroscopically determined redshift of 5.60 and was the subject of a separate publication (Weymann \etal\ 1998), and will not 
be discussed further here.

\noindent {\bf NIC645.0:} \\
A well-defined minimum exists at a redshift of 5.52, and there is no reasonable
secondary minimum. This is an extremely faint object, barely discernible in both
F110W and F160W and somewhat diffuse in F110W. The F814W image is readily apparent and slightly
elongated, (and hence has a WFPC2 identification), although there is no convincing
evidence for flux in the F606W image. In this respect this object is somewhat
similar to NIC184.0, (which has the same photometric redshift).

\noindent {\bf NIC586.0 and NIC107.0:} \\
We consider these two objects together since they are separated by only about 1.0
arcsec. Both objects are extremely faint. 
The deepest minimum for NIC586.0 yields the photometric redshift
of 5.68, but the minimum is not very well defined nor is the fit very good.
The same remarks apply to the $\chi^2$ curve vs. redshift for NIC107.0,
whose deepest minimum yields a photometric redshift of 5.92.
The F160W image of this pair shows one compact and one slightly diffuse
object in a field with several other nearby objects. There appears
to be no doubt about the reality of these two objects on the
basis of the F160W image. However, the flatfield properties of the F110W image
in this area of the chip are rather grainy and noisy and it is difficult
to compare the morphology of the two images, and the F110W photometry is
quite uncertain. There is a barely discernible very compact object in
the F814W image at the appropriate position for NIC586.0, but nothing visible
for NIC107.0. No convincing evidence for either object appears in the
V image.

\noindent {\bf NIC118.0 = WFPC2 4--601.0:} \\
This object was discussed in TWS. It is not a robust object in the sense described above,
but we discuss it because it has the highest photometric redshift in the
TWS table and because its interpretation is not at all clear.

Our measured 0.6" AB magnitudes  for the F160W, F110W, F814W, and F606W wavebands
are 27.79, 27.68, 30.41 and 30.44.  In a qualitative way,
this might be thought consistent with a high
redshift, and our formal
fit has a fairly well defined minimum at the photometric redshift of 6.56.
However at the photometric redshift given in TWS, it is difficult
to understand how there could be any flux whatsoever in the F606W band.
%%% RAY - start
The significance of the \mV\ and \mI\ detections in our convolved
images is only about $1\sigma$ but in the original WFPC2 images this object
is detected with a $5.75\sigma$ significance. Visual inspection of the images makes it
almost certain that there are positive detections in the \mV\ and \mI\ wavebands and that
the \mV\ flux
is not substantially less than the \mI\ flux (see discussion and figure in
the appendix).
%%% RAY - stop

Alternatively, the object might be a lower redshift, but highly reddened object.
A second alternative involves the chance superposition of two objects of quite
different redshifts. None of these interpretations is at all satisfactory, as
discussed in some detail in the Appendix.

We close the discussion on high redshift candidates 
by comparing our photometric redshifts of the 19 objects with 
redshifts $\ge 4.5$ with the only other
extensive compilation which extends to high redshifts in our field, namely the catalog
of Fernandez-Soto \etal\ (1999), hereafter FLY.
These authors provide a list of 170 photometric redshifts which fall in our deep NICMOS field
on chip 4, which is to be compared with the list of 282 in TWS. FLY based their photometric
redshift estimates on the 4 WFPC2 bands augmented by ground-based observations in the
standard J, H, and K bandpasses. These ground-based observations have, of course, much
lower spatial resolution than the NICMOS images, and the J and H observations do not go
nearly as deep as the F110W and F160W NICMOS fluxes. Consequently, for the very faint
objects on our list, these ground based observations generally do not yield statistically
significant detections. Nevertheless it is of interest to compare the results from the two
catalogs.

As explained in TWS we cannot
make unique associations in all cases with objects from the two lists because of differences
in the weight given to different bandpasses in composing the detection images, and in the
parameters used by SExtractor in breaking up significant pixels into separate objects. Nevertheless,
if we examine the histogram formed by looking at the distance between each object in the list
of each of the 170 objects in FLY to the nearest object in the TWS catalog, there is a
large cluster of objects with agreement in coordinates less than 0.15" whose median value is
about 0.05". There are 155 such objects. 
Of the remaining 15, all but one are either
so near the boundary that we rejected it from our list of 282 objects, or else they fall
within the boundaries of a clump of pixels which, for our SExtractor parameters 
applied to our images, are treated as part of a nearby object.
Evidently there are 127 objects in the TWS list which have no clear association 
with an object
for which FLY give a photometric redshift, though in many cases the objects are
obviously present but apparently too faint, given the data available 
to FLY, to derive photometric redshifts.

If we confine ourselves to objects in any one of these three lists with redshift
of 4.5 or greater, then  of the list of 15 FLY objects which have no TWS unambiguous
identification, there is one object (FLY \#393) which has a photometric redshift of 4.56.
On the F160W image it appears as an appendage on the very large low redshift galaxy NIC 26.100 and
was not separated with our SExtractor parameters. However, on the V and I images
it appears cleanly separated and is almost certainly a separate object.
In the list
of 127 objects in the TWS catalog not appearing in the FLY catalog there are 13 out of
our high redshift list of 19, so there is no basis for comparison among these objects:
NIC \#s 108, 1075, 150, 92, 693, 562, 96, 1040, 645, 586, 248, 107, and 
unfortunately, 118 = WFPC2 4--601 the object
discussed at length in the Appendix. Finally, among the 155 objects which can be unambiguously
associated in both catalogs, there are two in the FLY catalog which do not appear in our list
of 19 high redshift candidates: NIC 103 (WFPC2 4--625.1) and NIC 1062.0 (WFPC2 4--600.0 = FLY 471), however
in both cases the TWS redshift is just below our cutoff with photometric redshifts of 4.48, compared
to 4.52 given in FLY for NIC 103 and 6.52 for NIC 1062.0. We return to NIC 1062.0 below.

Our high redshift list contains two objects for one of which the discrepancy between FLY and TWS
is mild (NIC 131 = WFPC2 4--530 where TWS obtain z = 4.56 and FLY obtain z = 4.32) and one serious discrepancy
(the non-Robust object NIC 287 = WFPC2 4--148.0 where TWS obtain 4.64 and FLY obtain 1.24). 

In the case of NIC 287 = WFPC2 4--148.0 = FLY\#305, our fluxes give a much better fit at z = 4.64 than
at 1.24 in all three colors (V-I), (I-J), and (J-H).

The image of NIC 1062 = WFPC2 4--600 is very curious. It
lies about 0.7 arcsec from a much brighter object, NIC61.
It is most prominent in the F814W image, where
it appears as a somewhat diffuse object with  only moderate central concentration. 
The F606W image is very 
much weaker and the SNR is much lower than in the F814W image, and its flux
will depend upon exactly which pixels are included in the measurement.  FLY used the F814W
profile as a template for the aperture in which all the other bands were measured, and obtained
an AB V-I color of about 2.77--this is presumably what drove their very high redshift value of 6.52

The NICMOS F160W and F110W images are unexpectedly weak, and the contamination from NIC61
appears to be more severe than in the F606W and F814W bands. This makes the flux measurements
in the two NICMOS bands problematic. While in principle we might attempt
to deblend the NIC1062 and NIC61 images, the SNR of NIC1062 is not
high enough in the two NICMOS bands to make this practical. There were
12 pixels in the original ``detection map'' used in TWS for the NIC1062 fluxes. 
In an attempt to minimize the contamination from NIC61, we dropped three pixels
between NIC61 and NIC1062 and remeasured the fluxes with SExtractor. 
For these measurements we used only the flux contained within 
 a 3x3 pixel square centered on the coordinates of NIC1062 in the F814W image
rather than the 0.6" aperture fluxes of TWS.
Even in this case however,
the F160W flux and especially the F110W flux may suffer from significant 
contamination by NIC61. 

Nevertheless, taking at face value these new flux measurements we still obtain virtually the same
redshift for NIC1062 (4.64) compared with that given in TWS (4.48). The fit obtained predicts a 
flux in the F110W band which is about twice that observed, but the SNR in this band is only
about one so the discrepancy is not necessarily indicative of a failure in the template.

On the other hand, the  very high redshift of order 6.5 obtained by FLY seems very unlikely:
Our best fit at this redshift to the fluxes measured in the 3x3 patch
predicts an F814W flux which is nearly 7 sigma lower than the observed
value. The situation is even worse if, as suggested above, we have overestimated the F160W and
F110W fluxes because of the contamination from NIC61.

In summary, the detection of 9 robust high redshift candidates with five candidates
at redshifts greater than 5 in an area of less than $1.0^{\Box '}$
indicates that adding the near infrared bands to the deep optical
images is an excellent way to identify high redshift objects.  The flux
measurements in the F160W band  essentially double the range of
identification of high redshift galaxies via the strong Lyman break.
It is clear from Fig. 4 of TWS that these galaxies are at or near $L^*$ 
luminosities, indicating that such luminosities exist at high redshifts
where the hierarchical models expect very few high mass galaxies.
However, since these galaxies have SEDs that are dominated by very high
mass, extremely luminous stars, high luminosity does not necessarily
imply high mass.  In the next section we examine the question of small galaxies
at intermediate redshifts.

%%%% SECTION 3
\section{Selecting Compact, Intermediate  Redshift Galaxies}
\label{sect_selection}

As noted in the introduction, an exact one-to-one correspondence does not
exist between the list of
objects presented in this section and that which would result from a
similarly selected subset of objects in
Table 1 of TWS. 
As explained in
TWS the selection of objects for inclusion in that paper was based upon a
``$\chi^2$'' map following the prescription of Szalay, Connolly, and Szokoly (1999). 
There, since we had a
particular interest in identifying objects to the highest possible redshift, we
used only the equally weighted F814W,
F110W, and F160W  wavebands in composing the $\chi^2$ map. For the present purpose,
at the low end of the intermediate redshift regime in which we are interested, there
will be important information in both the F450W and F606W WFPC2 images, and hence we
used a $\chi^2$ map based upon equal weights for the F450W, F606W, F814W,
F110W, and F160W  wavebands. 
However, we used the same significance level to
define ``significant pixels'' (2.3) and the same required minimum
number of contiguous pixels (3) in the SExtractor algorithm 
(Bertin \& Arnouts\ 1996) as used in
TWS. Since most of these very small objects are at the very
edge of our detection threshold, we elected to measure ``isophotal'' fluxes (or, more
precisely, the flux contained in all the ``significant'' pixels as defined in
the $\chi^2$ map) rather than fluxes measured through an 0.6 arcsecond aperture, in
order to keep sky noise to a minimum. Otherwise, the methodology for selection, flux
measurement, flux error estimation, the estimates for the 
photometric redshift, template type and internal extinction as
well as their attendant errors,  are exactly as described in TWS.
The higher resolution images were convolved to match 
the F160W images. 
The goal was to select galaxies that are centrally concentrated 
as well as those where we are only detecting a bright knot measured
above our detection threshold. 

For our sample of compact objects, we then selected a subset of the
objects detected by SExtractor with the parameters described above, using 
the following criteria:
\begin{itemize}
\item  The ``isophotal'' area, as measured by SExtractor on the $\chi^2$ map, 
is $\le 0.2 ^{\Box ''}$. 
\item The photometric redshift is $1.0 \le z \le 4.5$.
\item The SExtractor parameter FLAGS=0 was required, so we only select
objects that are not blended/overlapping with others and are not
on the edge of the image.
\end{itemize}
This resulted in 93 selected galaxies.

As in TWS we tested the robustness of the photometric 
redshifts determined for this sample by perturbing the fluxes in 
each of the 6 wavebands, and then running the perturbed fluxes
through the photometric redshift estimator. 

This set of small compact objects tends to be at least as faint as
the high redshift candidates in the previous section, so that once again
the photometric errors are large and the resultant
uncertainties in the estimated photometric redshifts
are very large. In fact, of the 93 galaxies in our sample, only 11
have percentile values z$_{10}$ and z$_{90}$ within  
$\Delta z=0.5$ of the value determined from the unperturbed fluxes.

%%%% SECTION 4
\section{Averaging Images to Place Limits on Extended Flux}
\label{sect_average}

The 11 robust galaxies identified in the
previous section may be separated into three groups:
(i) Those with detectable F300W flux with photometric redshifts less than
$\sim2.0$, of which there are two. (ii) Those with detectable F450W flux
but little or no detectable F300W flux, whose photometric redshifts cluster
around $z \sim 2.7$, of which there are six. (iii) Three additional objects
with little or no detectable flux in either F300W or F450W whose redshifts
are around $z \sim 3.9$. Groups (ii) and (iii) we refer to as ``F300W dropouts'' and
``F450W dropouts''. 

For the remaining 82 galaxies, the
photometric uncertainties are too large at these faint flux
levels to obtain individual reliable photometric redshift
estimates. Instead, we simply visually inspected the images
in all six wavebands and placed them into one of the three
categories above. Even though these are faint objects the quality of
the HST images and the availability of six wavebands makes grouping
them into categories with and without F300W/F450W flux very straightforward.
While this procedure certainly does not
provide an accurate redshift discriminant it should provide
a rough guide to the redshift range appropriate for these
three groups.

From the two groups of F300W and F450W dropout galaxies thus
assembled we next selected those 
which appeared to be ``truly small''. This was done
by visual inspection of the images, keeping only those
which have no close companions and show no obvious extended flux. 

The objection might be raised that this is a circular
procedure, in which by selecting objects showing no extended
flux we find objects with no extended flux! But our objective
is to discover whether there is {\it any} subset of the
small galaxies which show no extended flux.
This procedure resulted in 14 ``isolated F450W dropouts'' and
23 ``isolated  F300W drop'' galaxies.
Only two of the galaxies with robust photometric redshifts are
included in this subsample. 
We then averaged together each waveband at the galaxy positions
and used the fluxes measured from these images
to obtain a `mean' photometric redshift for each `averaged' F300W drop
and F450W dropout galaxy.
The F300W  dropout mean redshift is in fact $z = 2.7$ and
the F450W dropout mean redshift is $z = 3.9$.
In both cases the hottest galaxy template
(a 50 Myr starburst; template \#6 in TWS)
with a small amount of reddening is preferred.
Images for all six HST wavebands for the (averaged) F300W  and F450W dropouts
are shown in Figure~\ref{f_ubmos}.  
\placefigure{f_ubmos}
Overplotted are 1.0 arcsecond diameter apertures. The galaxies
still appear compact in the averaged images.
The galaxies in the averaged samples have similar fluxes but we were concerned
that the brightest galaxies might dominate the average.  We 
did experiments removing the brightest one or two galaxies 
from the stack but this didn't change the results. 

Since we expect any older stellar population surrounding the small nucleus to
be redder than the population comprising the nucleus itself, one check for the
existence of such an older population is to 
measure the colors of the galaxies in successively larger
apertures.  
Within the uncertainties they are constant (Table~\ref{t_color}).
\placetable{t_color}
It is difficult to get accurate photometry at these faint magnitudes
so we measured the colors with both the SExtractor and IRAF photometry 
routines. These give consistent results. The errors calculated
for the photometry are of order $\pm$ 0.1. It can be seen from the 
table that the uncertainties are at least this large as the 2.5"
aperture gives a slightly fainter magnitude than the 1.5" aperture 
for the F160W filter. 
%The uncertainties in the photometry are at least $\pm$ 0.2
%magnitudes at these faint flux levels.
 
There is no evidence for diffuse,
redder flux outside of the core.  The F160W images for these
``averaged'' galaxies were remeasured with SExtractor and yield 
isophotal diameters of $\sim 0.5"$, which corresponds to
$\sim 4 kpc$ (H$_{0}$ = 65 km~s$^{-1}$~Mpc$^{-1}$, q$_{0}$ = 0.125).
If we measure the $3\sigma$ 
limiting depth in the blank sky at 1.6 microns in a 
$1.5" - 2.5"$ diameter annulus we obtain an F160W AB magnitude
of 29.3 for
the F300W  dropout and 29.4 for the F450W dropout.  
Limits on the amount of previous star formation that we can 
we infer from these limits are discussed in \S~\ref{sect_SF}.

%%%% SECTION 5
\section{The Expected Extended Flux from Compact Galaxies}
\label{sect_exp_flux}

In \S~\ref{sect_average} we set approximate empirical limits on the
count rate in the F160W band through an annulus with inner diameter $1.5"$ and
outer diameter of 2.5" associated with the mean of the isolated F300W  and
F450W dropout
galaxies described above.
Our failure to detect such extended flux
may be interpreted in at least two ways:
\begin{enumerate}
\item They are either the bright nuclei or star--forming knots in galaxies
with significantly lower average surface brightness made up of an older, cooler
population (we make no attempt to distinguish between these two
possibilities).
\item  They are young protogalaxy fragments which are not
embedded in an older stellar population.
\end{enumerate}
Colley \etal\ (1996) argue for the
former of these two possibilities on the basis of the two-point
correlation function, but our sample from the NICMOS field contains 
too few objects to make this a decisive test (see \S~\ref{sect_corr}).

When we repeat the Colley \etal\ analysis on the HDF north WFPC2 
chips 2,3 and 4 optical data set (see \S~\ref{sect_corr})we also find  a
significant signal in the autocorrelation function on small
scales, but visual inspection of the images shows that the
galaxy pairs responsible for the signal are often obviously
two pieces of the same galaxy. Clearly, in such cases, the Colley \etal\
interpretation of compact ``galaxies'' is the correct one. 

\subsection{Simulating Extended, Old Stellar Populations} 

Alternatively, we can ask whether surrounding lower surface
brightness material from a putative older population would be expected
to be detectable or not if it were placed at either of the two
representative redshifts ($z=2.7$ and $z=3.9$) inferred for the
F300W and F450W dropout subsamples described above, and what the
limits are for the stellar mass of an extended older population.
For illustrative purposes, and 
to stay as close as possible to the data set we are analyzing, we selected
a relatively bright galaxy from the HDF, WFPC2 4--378 (Williams \etal\ 1996),
with a photometric redshift of $z = 1.20$ (NIC 124.000 and 1022.00, TWS;
the redshift has since been measured
spectroscopically at z=1.225 by Cohen \etal\ 2000.)

This object provides a good example of a galaxy with
bright star--forming regions surrounded by an
older population or less actively star--forming regions.
To test the robustness of the photometric redshift estimate
with respect to where in this galaxy the flux is measured we
selected 4 different aperture positions:  1=bright F160W nucleus,
2=bright F814W off-nuclear knot, 3=dim F814W off-nuclear region,
and 4=the whole galaxy. The flux was measured in 0.6" diameter apertures
for positions 1 through 3 and a 2.6" aperture for position 4.
These are marked on the F606W+F814W image shown in
figure~\ref{f_photz}. 
\placefigure{f_photz}
It has been scaled and rotated
to match the NICMOS orientation and resolution
and the image stretch has been set to
emphasize the lumpy nature of the morphology.
The F606W$-$F160W colors range
from 1.91 for position 1 to 0.86 for position 4. The photometric
redshifts for each of these apertures were $z=1.20$ or $z=1.25$.

We then measured the flux in an 0.6---1.5 arcsecond annulus
centered on the bright nucleus. This is the region we would hope to
detect at higher redshift if the compact galaxies in our sample
are embedded in a comparable extended source.
The best least squares fit between the observed broadband {\it annular}
fluxes and those calculated on a grid of models in
redshift--reddening--population type space for this galaxy is very
good and constrains the spectral energy distribution over the range
covered from the F300W band to the F160W band fairly tightly
(even though there is
ambiguity between, {\it e.g.,} a very hot, but internally
reddened population and
a somewhat cooler but unreddened population, c.f. the discussion
in \S~6.3 of TWS).
With this best fit model,
and the observed flux in the bandpasses which most nearly correspond to
the 1.6 micron bandpass if the $z = 1.225$ galaxy were redshifted to
$z = 2.7$ or $z = 3.9$,
we can calculate the expected flux at higher redshift.
(In this calculation we ignore the change in proper
length with fixed angle, since over this redshift range this change is
very small for most cosmological models.)

Figure~\ref{f_compare} simulates what we would expect to 
see if our F300W  and F450W dropout galaxies were embedded 
in WFPC2 4--378.0 at a redshift of $z = 2.7$ or $z = 3.9$ and were observed  
at 1.6 microns.  
\placefigure{f_compare}
The redshift $z = 2.7$ case is on the left and 
the $z = 3.9$ case is on the right in the figure.
The top panels again show the averaged dropout galaxies at F160W.
The center panel shows how WFPC2 4--378.0 would 
look at 1.6 microns when scaled to the expected brightness 
at $z = 2.7$ and $z = 3.9$.  The circles overlaid on 
the images are 1 arcsecond in diameter.
To create the simulated higher redshift images we 
used the photometric redshift SED template that best fit 
the measured fluxes of the galaxy at $z=1.225$ and calculated 
what we would detect in the F160W bandpass for this galaxy at if it 
were at $z = 2.7$ or $z = 3.9$. 
We ignored any changes in angular size as
the angular size relation is roughly flat for $z > 1$.
We dimmed the F606W+F814W image showin in 
figure~\ref{f_photz}
appropriately and added this 
and added this to blank 
`sky' regions from the full NICMOS F160W image that have 
been averaged together.
The bottom panels were created by adding the dimmed versions of
4--378.0 to the dropout galaxies shown in the top panels. 
Though the full extent of 4--378.0 would not be visible at either
$z=2.7$ or $z = 3.9$, comparison of the top and bottom panels
makes it obvious
that the resulting simulated version of WFPC2 4--378
at these higher redshifts is larger than the isolated galaxies
we have detected, and the flux is readily measurable outside of a 
$0.6"$ aperture.

In \S~\ref{sect_average} we made averaged images that were centered on 
the compact source in each field. If the objects in our sample are truly
off-nuclear star--forming knots in an underlying star--forming galaxy, then
to make
a simulation that more accurately mimics what these averaged images might 
contain, we have again taken the galaxy WFPC2 4--378.0 and randomly rotated 
the galaxy around 2 different bright knots, marked {\bf A} and {\bf B}
in  figure~\ref{f_knots} which shows the same image as figure~\ref{f_photz}.
\placefigure{f_knots}
We then take 20 of the randomly rotated galaxies and average 
these together in the same way we made the F300W  and F450W 
averaged dropout galaxy images. 
The result is illustrated in figure~\ref{f_rotgals}. 
\placefigure{f_rotgals}

Both of these `averaged' galaxies now show much more
diffuse morphology. (Note that though WFPC2 4--378.0 has a bright nucleus in the F160W
band, it is not apparent in the F606W and F814W images.  Therefore when we
make the dimmed, averaged images for z=2.7 and z=3.9 centered on the
off-nuclear knots using the HDF optical images, we are not smearing a
bright nucleus into appearing as extended flux.)
The results from adding these simulated galaxies to the
F300W dropout galaxy and F450W dropout
galaxy give a result very similar to what we saw in
figure~\ref{f_compare}.

At $z=2.7$ the diffuse flux is easily detected by eye in
the images.  At $z=3.9$ it is much less obvious. However, when
the magnitudes are measured in successively larger
apertures, the extended flux
is definitely measurable. In Table~\ref{t_color} we showed that the
F160W magnitudes for the F300W dropout and F450W dropout galaxies
didn't change when measured in 0.6", 1.5", and 2.5"
apertures. After adding the simulated, extended galaxies
to these images, the F300W dropout galaxy ($z = 2.7$) is $\sim$0.4 mags brighter
in the 1.5" aperture and $\sim$0.9 mags brighter in the
2.5" aperture in both the knot-A and knot-B cases.
The B-drop galaxy ($z = 3.9$) is $\sim$0.9 mags brighter
in the 1.5" aperture and $\sim$1.5 mags brighter in the
2.5" aperture. Though faint, the flux from the extended galaxies is
detected over an extent of 2 arcseconds.  If the
compact galaxies we have estimated to be at redshifts
z $>$ 1 were embedded in more diffuse, larger galaxies,
such as WFPC2 4--378, we would be able to measure this extended flux
in the NICMOS images.

\subsection{Limits on The Star Formation History of the F300W  and F450W Dropout Galaxies}
\label{sect_SF}
While the foregoing examples suggest that significant star formation occurring outside
the nucleus of distant galaxies, similar in character to WFPC2 4--378 would be detectable
in our averaged dropout galaxies, we now address the more quantitative question of
limits on star forming activity in these dropout galaxies.

In \S~\ref{sect_average} we measured the  
$3\sigma$ upper limit on the flux in the F160W filter in an annulus
of $1.5" - 2.5"$ diameter for the averaged F300W dropout
and F450W dropout galaxies.
 These measurements give F160W=29.3 for
the F300W dropout and F160W=29.4 for the F450W dropout. This
annulus corresponds to an inner and outer radius of about 6 and 10 kpc,
respectively---an annulus centered roughly on
a strip whose center is comparable to the sun's galactocentric
distance.  We now investigate what limits on the star formation
history in this region the flux limits imply for these two averaged galaxies.

%Given what is known about both the number density and redshifts of the highest redshift
%galaxies and QSOs, it appears that very little star formation
%occurred at redshifts much beyond six. 
In the following we consider
models for star formation which commenced at about $z \approx 6.0$
Although we have referred previously to a possible underlying ``old'' stellar population,
the intervals of time between the epochs corresponding to $z \approx 6.0$ and z=2.7 (for
the F300W dropouts) and $z \approx 6.0$ and z = 3.9 (for the F450W dropouts) are only about
1.4 and 0.6 Gyrs, respectively, so that we would expect
an intermediate age, rather than old stellar population to be present. 

Out of the infinite number of possible star formation histories, we consider three:

(i) An initial burst of star formation of duration 50my which began at a
    redshift slightly less than 6.3 and ended at a redshift of 6.0
(ii) An intermediate case in which star formation began at z = 6.0, decayed
      exponentially with a time constant of 1 Gyr, but with a constant amount
      subtracted from this exponentially decaying star formation rate, so that
      at the epoch of observation the star formation rate had reached zero.
(iii) A burst of star formation of duration 50mY which ended just at the epoch of
      observation.
The models include dust extinction using the procedure described
in TWS, section 3.2.  The best fit includes a small amount of
extinction.
The results are shown in table~\ref{t_masslimits}. 
The three rows represent cases (i), (ii) and
(iii) while columns 2 and 4 give the total mass of stars formed 
over the observed annulus which would just produce
the observed 3$\sigma$ limit on the F160W flux in the annulus.
Columns 3 and 5 give the corresponding instantaneous
rates of star formation (in solar masses per year) for case (iii).

\placetable{t_masslimits} 

In the rest frame of objects
at z = 2.7 and z = 3.9, the center of the F160W band pass at 1.6 microns occurs
at 4324 and 3265 \AA\ respectively. The limit for the F300W dropout at z = 2.7 is
of course considerably more stringent, simply by virtue of being closer than
the F450W dropout. For case (i), which postulates a single initial brief burst
of star formation at z = 6.0, occurring anywhere in the outer annulus, the
hottest stars have left the main sequence. For the F300W dropout case the
F160W band flux samples the region just longward of the Balmer discontinuity
and a spectrum would show strong Balmer lines from an intermediate age
population. For the intermediate case (ii) the SED is qualitatively similar, but
a more extended period of star formation implies that the limit on the total mass of stars
formed is significantly less.  Case (iii) does not refer to an older population, but merely
says that the star formation responsible for the images actually seen in the
HDF was confined to a small region, and very little could have occurred in
the surrounding annulus. The situation for the F450W dropouts is qualitatively
similar, except that in addition to the fainter flux as a consequence
of the greater distance, the F160W band for this redshift samples rest
wavelengths which are shortward of a strong Balmer discontinuity,
though in case (ii), the interval of time between the start of the
star formation at a redshift of 6 and the epoch corresponding to
the redshift of z = 3.9 is smaller than the decay time, and consequently
there is still significant flux from fairly hot stars.

%\subsection{Implications Relative to the Hierarchical Model of Galaxy 
%Formation}
%
%Many current theories of galaxy formation invoke a hierarchical model
%in which present day galaxies are assembled via numerous mergers of
%lower mass galaxies.  An obvious question is whether the galaxies
%discussed in this section are candidates for such earlier, lower mass
%objects.  The answer is probably ``yes'' since they exist at early
%times, are smaller than many present day galaxies, and lack extended
%regions in which significant star formation has occurred.  However, size
%predictions for hierarchical formation (\eg\ Rauch, Haehnelt \& Steinmetz 1997) 
%predict
%that the young stellar population is distributed in an area of diameter
%5 kpc or less for the redshift range we are considering.  Our limits on
%size of 10-15 kpc diameter is consistent with these sizes but does not
%directly probe the predicted size ranges.

%%%% SECTION 6

\section{Spatial Galaxy Correlations}
\label{sect_corr}

\subsection{Galaxies Selected from WFPC2 Data}
We examine here the scale on which the
galaxies are correlated in the WFPC2 fields to compare with 
the results of Colley \etal\ (1996). They found a positive
signal in the correlation function on  
scales $\le$ 1 arcsecond for high and low redshift 
color cuts.
In particular, they found that for the smallest objects,
as well as 
those with colors that suggest they are at redshifts z $>$ 2.5,
there is a strong autocorrelation for
scales $\le 0.5$ arcseconds. The color cut used was
F300W -- F450W $>$ 1.2 $+$ F450W -- (F814W $+$ F606W)$/2$
(Steidel \etal\ 1996). 

As an estimator for the two-point angular correlation function,
$w(r)$, we used  
\begin{equation}
w(r) = { N_{dd}(r) \over  N_{dr}(r) }
{ 2n_r \over (n_d - 1) } - 1
\label{equ1}
\end{equation}
(Efstathiou \etal\ 1991; Roche \etal\ 1996), where $N_{dd}$ is the 
number of pairs in our data set, $ N_{dr}$ is the number
of pairs in $n_r$ randomly placed objects in our field, and 
$n_d$ is the number of objects in our data set. 
We calculated $w(r)$ in 0.25 arcsecond bins 
out to a radius of 5.0 arcseconds with $n_r$ = 10,000. 

We extracted galaxies
from combined F606W + F814W images of the WFPC2 chips 2, 3, and 4 
HDF north fields with SExtractor
(2.5$\sigma$ threshold, 10 pixel minimum area).
We next selected a catalog of galaxies 
matching the above color criterion
from $71.0" \times 69.2"$ central regions of the chips
that miss the lower signal-to-noise regions near the edges.
This results in 273, 215, and 222 galaxies from chips 2, 3, and 
4 respectively.  
The results for the autocorrelation function of this high redshift sample
are shown in figure~\ref{f_wfpccut}.
\placefigure{f_wfpccut}
The data are plotted as 
stars (WF2), triangles (WF3), and circles (WF4). The $\pm 1\sigma$ range
for no correlation, determined from 
500 realizations of 10000 randomly placed points, is shown 
as the crosshatched region around $w(r)=0$.  
The high redshift color sample has a strong signal 
for all three WFPC2 chips on scales less than 0.5$"$, 
confirming the result found by Colley \etal\ (1996). 
The signal is significant at the $3-7\sigma$ level, and
peaks at $0.25" - 0.5"$.  No significant autocorrelation
is seen at separations $> 1"$. 
When we examine the pairs on 
the F606W image, it is clear that SExtractor has counted
2 pieces of what appear to be the same galaxy in many of these
cases. Two examples of this are WFPC2 4-603.0 (NIC110.000) and
WFPC2 4--555.1 (NIC141.112), also known as the ``hot dog''.
Their respective redshifts are z=2.56 (photometric, TWS) and
z=2.803 (spectroscopic, Steidel \etal\ 1996).

\subsection{Galaxies Selected from NICMOS Data}

We ran the same analysis discussed in the previous section
on our catalog of 93 small galaxies 
from the NICMOS data set. To determine the uncertainty we ran the 
same estimator on 500 sets of 93 randomly placed galaxies.  The results 
are shown in figure~\ref{f_small}.
In contrast to the results discussed above for the WFPC2 data, we see 
no correlation signal in the NICMOS data.

\placefigure{f_small}  

It is difficult to interpret whether this result means that 
(a) the small objects are not correlated, or (b) the lower resolution
in the NICMOS image compared to the WFPC2 data and/or our smaller
survey region preclude measuring 
any signal above that expected from random superpositions.
In the near-infrared many galaxies will appear smoother since we 
are observing older stellar populations and SExtractor 
will be less likely to break them into multiple pieces.
To determine how the correlation signal would be affected by
examining a smaller region of the sky, 
we then reran the analysis of the WFPC2 color-selected catalog,
but just used the data from $0.65^{\Box '}$ regions in each
chip, the areal size covered by our NICMOS observations. 
The results are shown in figure~\ref{f_wfnareacolor}.

\placefigure{f_wfnareacolor}

Using the smaller region the correlation signal is stronger in WF4, 
about the same in WF2, and has disappeared in WF3 when
compared to the full area measurements shown in figure~\ref{f_wfpccut}. 
Our $0.65^{\Box '}$ region may 
hinder the detection of a real signal with these faint objects,
but it may also be that our small galaxies constitute a set of objects
with genuinely different correlation properties than the set defined in the
Colley \etal\ (1996) analysis.
Thus, we can neither confirm nor reject the the hypothesis
that the 93 small galaxies from the NICMOS data set are
uncorrelated.

In summary, we confirm the Colley \etal\ (1996) result finding a
positive signal in the autocorrelation function on
scales $\le$ 1 arcsecond for objects detected in the WFPC2 images 
with colors suggesting they are 
at high redshift.  
We do not detect any significant signal 
in the autocorrelation function for the small, 
moderate to high redshift galaxies detected in the NICMOS 
image.  However, we cannot make a decisive interpretation of this 
due to the small size of our sample. 

\subsection{Correlation of Blue and Red Galaxies}
\label{sect_bluered}

It was apparent just by looking at the optical and 
infrared images that there were a number of cases
where faint blue smudges were next to brighter red galaxies
but were not obviously pieces of the same galaxy.
An example is shown in figure~\ref{f_bluereddetail}.
The left panel shows a $4.5" 
\times 4.5"$ WFPC2 F606W image and the 
right panel shows the same region through the F160W filter. 
The galaxy just left of the center in both images is
WFPC2 4--307.0 (NIC166.000).  WFPC2 4--312.0 is the
galaxy $1.5"$ to the right of WFPC2 4--307.0 and WFPC2 4--279.0 is
the galaxy $2.1"$ above 4-307.0 in the F606W image.
The two blue objects, WFPC2 4--312.0 and WFPC2 4--279.0, were
not formally detected by SExtractor using the S/N thresholds we
have adopted and thus do not have NICMOS identifications. However, 
measuring the F160W flux at the centroids expected yields AB F606W - F160W
color indices of $\le-0.3$ and $\le-1.0$ respectively
Conversely, WFPC2 4--307.0 is one of the brighter
objects in the NICMOS catalog and yields a color index
of $+4.2$. (See the figure caption for details). 
We estimate a photometric redshift of $z \approx 1.85$
for WFPC2 4--307.0. Though
WFPC2 4--312.0 and 4--279.0 could be galaxies unrelated
to WFPC2 4--307.0, their proximity
and color are also consistent with star--forming
regions 12-16 kpc from the center of WFPC2 4--307.0.

\placefigure{f_bluereddetail}

Pairs of red and blue galaxies such as these motivated us
to look for a statistical correlation between the very blue
and redder populations. 
Similar to the autocorrelation analysis described above,
we measured the separations of blue and red galaxies
with respect to what we might see with randomly placed samples
in our region.  We selected samples of galaxies with `blue'
($F606W - F160W \le 0.3$) and `red' ($F606W - F160W \ge 1$) colors then measured the
separations of all the pairs.   Here we might have expected to see more
pairs at separations of 1--2
arcseconds but no significant signal is detected.
As in the autocorrelation analysis though, we cannot use this
to completely rule out correlations that seemed apparent when
examining the images by eye due to the very small region surveyed.

%%% SECTION 7
\section{Discussion and Summary: Small Galaxies}
\label{sect_discuss}

We examined averaged images of faint ({F606W magnitudes $\sim$ 27--29), 
compact objects to search for extended, surrounding flux from older, 
fainter populations of stars.  We also simulated what extended galaxies, 
easily detected at z $\sim$ 1, would look like at higher redshifts
when measured in the infrared wavebands. We have determined that
though these galaxies would of course appear substantially fainter, the more
diffuse parts of the galaxies would still be clearly detected.
Our results suggest that at least some of the small, (isophotal area $\le 0.2 ^{\Box ''}$),
high redshift, ($1 \le z \le 4.5$),
galaxies detected in the HDF are bonafide compact objects, not
bright nuclei or star--forming knots embedded in
more extended stellar populations. The limits on the
flux in an annulus between 1.5 and 2.5 arcsec in diameter correspond
to limits on the total mass of stars which could have been formed
in this annulus of a few times $10^8\Msun$ for the U-drop galaxies
and slightly more for the B-drop galaxies, the exact mass limit
depending upon the star formation history assumed. Our simulations have shown
that a prototypical extended galaxy with bright clumps detected easily
in the HDF at redshift $z \sim 1$ (specifically WFPC2 4--378) would still be detected at $z = 3.9$
in the NICMOS observations at 1.6 microns. This
object has had substantial star formation extending to a radius of $\sim$10kpc. In
contrast, the objects forming our averaged U-drop and B-drop images are
consistent with being young protogalaxy fragments.  Our photometric
redshift estimator prefers the hottest galaxy template
(a 50 Myr starburst) with a small amount of reddening for both the
F300W and F450W dropouts, and gives mean redshifts of $z = 2.7$ and
$z = 3.9$ for these, respectively.

It has been suggested in previous work (e.g. Colley \etal) that compact
sources in the WFPC2 HDF images are subgalactic components at
redshifts $z > 0.5$ since they are correlated on scales less than $1"$,
corresponding to physical scales of less than 8 kpc
(\H0 = 65 \ksM, \q0 = 0.125).
Using the data from all the WFPC2 chips we confirm these correlations 
in objects selected from the optical images with high redshift colors.
However, we do not detect the correlation of close pairs of galaxies on 
small scales in the $\sim 0.65^{\Box '}$ region of the HDF that we surveyed with NICMOS.
This is likely due to one or more of the following factors:  
1) The area we surveyed with NICMOS is substantially
smaller than that covered by the WFPC2 observations. 
When we sample $0.65^{\Box '}$ regions of the WFPC2 images, 
where the correlation is found at a 
statistically significant level in all three WF chips, 
we no longer detect any correlation 
signal in the WF3 chip, simply due to the smaller area surveyed. 
2) The resolution in the NICMOS images is slightly lower
than the WFPC2 images, and in the near-infrared many galaxies
will appear smoother since we are observing older stellar populations 
and they will be less likely to be broken up into multiple pieces by 
the galaxy detection algorithm. 3) Our selection criteria for small galaxies
has yielded a set of objects whose correlation properties are in fact
different from those selected by Colley \etal.

When we visually examine examples of correlated pairs, many are 
clearly pieces of one galaxy.
Thus, the distinction between ``separate correlated
galaxies'' and patches of the same galaxy is rather artificial
and depends not only on the signal-to-noise ratio of the data
but also the parameters used in the source extraction algorithm.
This does not contradict our previous conclusion that 
there exist some isolated, compact objects which are not embedded in 
larger galaxies. 
What we are suggesting is that these correlations should be interpreted
with care and that many intermediate and high redshift compact galaxies 
do not have close companions nor extended flux surrounding them. 

Numerical simulations 
have shown that the progenitors of a galaxy at $z = 0$ formed in a hierarchical
clustering scenario would be detected at $z = 3$ as several
protogalactic clumps covering an area on
the sky the size of a WFPC2 chip or larger, and thus not
highly correlated (Haehnelt, Steinmetz \& Rauch 1996 [p. L96]; 
Rauch, Haehnelt \& Steinmetz 1997; Steinmetz 1998).  Our results are
consistent with, but do not confirm, this picture.

\acknowledgments

This research was supported in part by NASA grant
NAG 5-3042, which is gratefully acknowledged. We thank the 
anonymous referee for suggestions that improved the manuscript. 

\newpage

\appendix
\section{THE NATURE OF NIC118.0 = WFPC2 4--601.0}

As discussed briefly in \S2, the highest formal photometric redshift derived by
TWS in their catalog is for the object NIC118.0 = WFPC2-601.0 at a redshift of 6.56.

For simplicity of notation
we refer to the four wavebands F160W, F110W, F814W, and F606W 
as \mH, \mJ, \mI, and \mV, though 
readers should note that the HST wavebands are not the standard 
Johnson filters, particularly the F110W.
The associated color indices for our measured AB magnitudes in these four
wavebands through an 0.6" aperture of  27.79, 27.68, 30.41 and 30.44 yield
associated color
indices of \mJ\ -\mH\  = -0.11, \mI\ -\mJ\ = 2.73, and \mV\ -\mI\ = 0.03. It is
the latter two color indices which make the interpretation especially puzzling.
%%% RAY - start
Evidently, the precision of the \mI\ and \mV\ magnitudes and the associated color 
index is poor, though the agreement of our flux with that
given in the WFPC-2 catalog (Williams \etal\ 1996) is fairly
good. The significance of the  \mV\ and \mI\ detections in our convolved
images is only about $1\sigma$ but in the original WFPC2 images this object
is detected with a $5.75\sigma$ significance. Visual inspection of the images makes it 
almost certain that there are positive detections in the \mV\ and \mI\ wavebands and that 
the \mV\ flux
is not substantially less than the \mI\ flux. The original F606W and F814W WFPC2 images 
are shown in figure~\ref{f_append1}.  
%%% RAY - stop 
Inspection and measurement of the
profiles in the \mH\ and \mJ\ images clearly show the object to be definitely
non-stellar and somewhat irregular in appearance. It also appears to be extended
in the \mI\ and \mV\ bands, though in these two bands the flux is so weak that we have
not attempted to measure the profile. We therefore assume in the following that the
object is extragalactic.

\placefigure{f_append1}

As noted in \S 2, though the
fit has a fairly well defined minimum at the photometric redshift of 6.56,
attenuation by Lyman absorption would be expected to produce a much
larger \mV\ - \mI\ index. To explore this further and to see whether a 
somewhat smaller redshift might be compatible with the above flux measurements
we consider models for Lyman attenuation at 5 fiducial redshifts: 6.50, 6.25,
6.00, 5.75 and 5.50.  Table~\ref{t_a1}  shows the expected observed wavelengths 
of Lyman-alpha, 
Lyman-beta and the Lyman continuum for these 5 redshifts.

\placetable{t_a1}

The recent discovery of very high redshift QSOs from the Sloan Survey
(Fan \etal\ 2001) has allowed estimates of the Lyman attenuation to be extended
beyond the range where the empirical estimates of Madau (1995),  
were made. Extrapolation of the Madau results were used in calculating the TWS
photometric redshifts [Madau 1995 $+$ detailed tables provided as  private communiction to R. Thompson)

In particular, Becker \etal\ (2001) have discussed
the Lyman forest attenuation up to the highest redshift QSO which they
discovered (z = 6.28). 
Their results show that the region slightly
shortward of the Lyman-alpha emission is black to within the errors
of their measurements. They also set very low limits on the amount
of flux transmitted below the Lyman-beta emission line, but there
appears to be some flux between 7500-8000A, though one cannot be sure
to what extent this represents sky subtraction errors. 

Becker \etal\
model the expected flux at the Lyman-beta line by assuming that
the Gunn Peterson trough optical depths at Lyman-alpha and Lyman-beta 
scale as the ratio of the f-values (actually they should scale
as the ratio of the product of the wavelength and f-value), so in this
simple model the ratio of the Gunn Peterson trough at Ly$\alpha$ to that at
Ly beta (for a given redshift) should have a ratio of 6.24. 
This same modeling can be extended in a straightforward fashion
to include the Lyman continuum attenuation.  We refer to this
simple Gunn Peterson model as the ``uniform'' model.

In fact,
the empirically-derived ratio of the attenuation from
Lyman-alpha to that associated with Lyman-beta (at a fixed absorption
redshift, not a fixed observed wavelength) derived by
Madau (1995) at somewhat lower redshifts yields
effective optical depth ratios (Ly$\alpha$/Ly$\beta$) which are substantially
lower--of order 2.0. 

The reason for this discrepancy must almost certainly be
due to the lumpiness of the gas (``clouds") which presumably persists to higher
redshifts. We have therefore used, as an
alternative to the ``uniform'' model above,
a  simple ``picket-fence'' model in which the Lyman-alpha optical
depth varies over small regions of redshift space between
a relatively small value ($\tau_{min}$) 
and a large value ($\tau_{max}$) with the
large value covering a fraction, f, of a given redshift interval. The
higher extinction in the $\tau_{max}$ component causes Lyman-beta
to be saturated in the region of redshift space covered
by this component. Unless sufficiently high SNR and high resolution observations
are carried out, the measured transmission will be the average of the
transmission associated with these two components. Additionally, if the high effective
optical depth of the Lyman-beta transmission at
very high redshifts  estimated by Becker \etal\ (for example 
$\sim 3$,) is interpreted in terms of the uniform model,
it requires a Lyman-alpha optical depth of $\sim19$, which
seems excessive given their published estimates, unless there
is a truly abrupt and huge discontinuity in this quantity.
We have therefore used combinations
of $\tau_{min},\tau_{max}$, and f whose properties are
in reasonable accord with these high redshift
QSO observations, but fixing them so that the effective ratio of Ly$\alpha$ to 
Ly$\beta$ optical depths is around 2, in accordance with the Madau (1995) 
results. With the run of these three parameters as a function
of redshift, the contribution from the higher Lyman lines and
continuum is determined. This may or may not
be in conflict with the limit on the Lyman-beta flux
observed by Becker \etal\ at the highest redshfits.
In this model,
one of the three parameters above is free and we have arbitrarily
fixed the thick component at an optical depth of 30.0. 

We show the effective
low resolution optical depths for both the uniform and picket fence
models for an emission redshift of 6.0 in figure~\ref{f_append2}.

\placefigure{f_append2}

There is
no difference in the two models in the region to the red of the
Lyman-beta absorption, but to the blue of it, because of 
the rapid drop in the ratio 
of the Gunn-Peterson optical depths with increasing
Lyman line number at a fixed redshift, the increasing number 
of Lyman lines capable of producing absorption does not compensate for 
the decrease 
in the Lyman-alpha optical depth with decreasing
redshift. In the picket fence model, the opposite is true, since
the drop in the effective opacity as one progresses to higher
Lyman lines is much slower, and in fact the curve for the
picket fence absorption resembles the empirical Madau curves
for lower redshifts, as expected.

Even this picket fence model probably underestimates the actual
attenuation shortward of the Lyman limit, since it is highly likely
that much thicker clouds intervene. Measurements at lower
redshifts (Giallongo \etal\ 2002) strongly suggest
that almost no Lyman continuum radiation is escaping.
However, it is conceivable
that a very luminous burst of hot stars in a protogalaxy could produce
its own proximity effect and allow some Lyman continuum radiation to escape.

To get the smallest V-I color in the face of the Lyman attenuation, we use a
very hot model, namely a burst of star formation lasting just 25 mY and ending
at the 5 fiducial redshifts, and with no internal or external dust reddening.
Using these two  simple models for the Lyman attenuation
produces the results shown in table~\ref{t_a2}.

\placetable{t_a2}

Note that the progression in the V-I color index for the Uniform model is not
monotonic with increasing emission line redshift because the reduction in the
I-J color  due to Lyman-alpha cutting into the I band reduces the I band
flux relative to the more-or-less unaffected J-band flux more rapidly
than the relatively weak Lyman continuum attenuation affects the V-band flux,
(though the V-H index is monotonic).  The strong attenuation in the Lyman continuum
in the picket fence model causes the V-I color index to strongly rise with
increasing redshift. In any case, it is clear that even the Uniform model cannot
come close to reproducing the observed colors, and the (in our opinion more realistic)
picket fence model fails even more badly. 

In fact, the attempt to find a solution
at very high redshifts without {\it any} Lyman attenuation fails: The reason is that
the intrinsic Lyman continuum edge in even the very hot 25mY model in the intrinsic
stellar population gives a V-I color index which is much too red, as shown in the
last model ``No Ly Abs".

We conclude that if subsequent observations {\it do} establish the very high redshift
nature of this object then it must have some extraordinary properties.

The alternative of a lower redshift object suffering from internal reddening is
no more successful: A shallow secondary minimum in the $\chi^2$ search
for the best combination of redshift, internal reddening and population template occurs
at a redshift of 0.2, while the tertiary minimum occurs at a redshift of 1.05. The
resulting color indices for $z = 0.2$ are 1.36 and 1.22 for V-I and I-J respectively,
while for $z = 1.05$ they are 1.22 and 1.16. 
Both of these sets of indices are very far from the observed values. The problem
is obvious: An amount of reddening sufficient to produce a very red observed I-J index
would also produce an even larger V-I index since the extinction increases with decreasing
rest wavelength.

\subsection{Superposition of Two Objects}

It is conceivable that NIC118.0 is actually a superposition of two objects with very different
redshifts. One would be a truly very high redshift object for which the Lyman attenuation can
produce the large I-J color, with a low redshift, hot dwarf galaxy providing some of the V flux.
Given the excellent positional agreement between the WFPC-2 and NICMOS images, together with
the intrinsic very rare nature of such a very high redshift object (and moderately rare
low redshift hot dwarf) such a solution seems to us quite artificial.

\subsection{Concluding Comments}

Although there are important differences, the dilemma posed by this object is reminiscent
of that posed by the object HDF-N J123656.3+621322 (Dickinson \etal\ 2000), which has
strong flux in $K_s$ and H, but no convincing detection in any of the shorter wavebands.
Its nature is still (as far as we are aware) uncertain, but these authors raise the
possibility (among others) of an AGN, with emission lines possibly influencing the colors.
Without further observations of NIC118.0, however, further speculation seems fruitless.

An attempt is being made by Spinrad \etal\ to obtain spectra of this object, but the results
are inconclusive so far.

The GOODS HST Treasury program will, unfortunately, probably not go deep enough in V or I to
detect this object, but the new Z-band observations may add an important clue. It may be
up to either NGST or the next generation of very large ground based telescopes to reveal
its nature.

%%% FIGURES \& TABLES

\clearpage
%%% TABLE 1: HIZ CANDIDATES
\begin{deluxetable}{llccclccccr}
\tablenum{1}
\tabletypesize{\scriptsize}
\tablewidth{0pc}
\tablecaption{High Redshift Candidates}
\tablehead{
\colhead{(1)} & \colhead{(2)} & \colhead{(3)} & \colhead{(4)} & \colhead{(5)} & 
\colhead{(6)} & \colhead{(7)} & \colhead{(8)} & \colhead{(9)} & \colhead{(10)} & 
\colhead{(11)} \\
\colhead{NICMOS} & \colhead{WFPC2} & \multicolumn{2}{c}{AB (F160W)} & 
\colhead{RA} & \colhead{Dec} &  \colhead{z$_{phot}$} & \colhead{Robust} &
\colhead{z$_{phot}$} & \colhead{z$_{phot}$} & \colhead{Rank} \\
\colhead{ID\tablenotemark{a}} & \colhead{ID\tablenotemark{b}} & \colhead{Total} & \colhead{0.6" ap.} &
\colhead{+12h36m }& \colhead{+62d} & \colhead{(best)}& \colhead{Flag} & 
\colhead{(5\%)} & \colhead{(90\%)} &  
}
\startdata
131.000 & 4-530.0 & 27.5 & 27.8 & 45.15 & 11:59.7 & 4.56 & R & 4.32 & 4.64 & 3 \\
1081.00 & 4-748.0 & 28.3 & 28.4 & 41.96 & 12:09.1 & 4.64 & NR & 0.48 & 4.88 & 13 \\
287.000 & 4-148.0 & 27.9 & 28.6 & 48.92 & 12:16.7 & 4.64 & NR & 0.48 & 4.80 & 11 \\
1075.00 & 4-526.0 & 28.6 & 28.8 & 45.11 & 12:00.4 & 4.72 & NR & 0.48 & 4.88 & 29 \\
274.000 & 4-200.0 & 26.0 & 26.5 & 48.37 & 12:17.3 & 4.72 & R & 4.48 & 4.80 & 2 \\
267.000 & 4-314.0 & 27.3 & 27.5 & 48.00 & 12:00.8 & 4.80 & R & 4.64 & 4.80 & 1 \\
150.000 & \nodata & 26.9 & 27.5 & 45.42 & 12:02.2 & 4.80 & R & 4.56 & 5.04 & 1 \\
96.0000 & \nodata & 27.4 & 27.9 & 44.54 & 12:36.1 & 4.88 & NR & 0.96 & 5.44 & 11 \\
693.000 & \nodata & 28.2 & 28.6 & 48.86 & 12:16.8 & 4.88 & NR & 0.96 & 5.12 & 12 \\
92.0000 & 4-499.0 & 27.4 & 27.7 & 44.44 & 12:17.2 & 4.88 & NR & 0.96 & 5.52 & 30 \\
562.000 & 4-663.0 & 28.8 & 28.9 & 43.91 & 11:54.4 & 4.88 & NR & 0.96 & 5.20 & 16 \\
277.212 & 4-169.0 & 24.6 & 25.1 & 48.71 & 12:16.7 & 5.04 & R & 4.80 & 5.12 & 1 \\
1040.00 & \nodata & 28.1 & 28.7 & 47.83 & 12:04.5 & 5.44 & NR & 0.00 & 5.76 & 17 \\
184.000 & 4-473.0 & 26.6 & 26.9 & 45.88 & 11:58.2 & 5.52 & SC & 5.20 & 5.52 & 1 \\
645.000 & 4-262.2 & 28.1 & 28.2 & 46.45 & 12:37.5 & 5.52 & R & 4.88 & 5.76 & 2 \\
586.000 & \nodata & 27.5 & 28.0 & 44.71 & 12:20.0 & 5.68 & R & 4.80 & 6.00 & 1 \\
248.000 & \nodata & 27.2 & 27.7 & 47.46 & 11:59.9 & 5.76 & NR & 0.08 & 6.00 & 19 \\
107.000 & \nodata & 27.1 & 27.7 & 44.72 & 12:18.8 & 5.92 & R & 4.80 & 6.32 & 2 \\
118.000 & 4-601.0 & 27.5 & 27.7 & 44.90 & 11:50.3 & 6.56 & NR & 0.24 & 6.88 & 20 \\
\enddata
\label{t_hiz_tab}
\tablerefs{(a) Thompson, Weymann, \& Storrie-Lombardi 2001; (b) Williams \etal\ 1996 }
\end{deluxetable}

\clearpage
%% TABLE 2
\begin{deluxetable}{lccccc}
\tablenum{2}
\tablewidth{0pc}
\tablecaption{F814W - F160W Colors for Averaged Galaxies}
\tablehead{
\colhead{Galaxy} & \colhead{Filter} & \multicolumn{4}{c}{Photometry Aperture}  \\
& & \colhead{Isophotal}  & \colhead{0.6"} & \colhead{1.5"} & \colhead{2.5"}
}
\startdata 
F300W dropout & F814W & 28.3 & 28.5 & 28.3 & 28.5 \\
              & F160W & 27.6 & 27.9 & 27.6 & 27.7 \\ 
              & F814W - F160W & 0.7 & 0.6 & 0.7 & 0.8 \\ 
F450W dropout & F814W & 28.5 & 28.6 & 28.7 & 29.0 \\
              & F160W & 28.4 & 28.5 & 28.8 & 28.7 \\ 
              & F814W - F160W & 0.1 & 0.1 & -0.1 & -0.3 \\ 
\enddata
\label{t_color}
\tablecomments{The apertures sizes listed are diameters. The 
calculated errors 
for the photometry are of order $\pm$ 0.1. It can be seen from the  
table that the uncertainties are at least this large as the 2.5"
aperture gives a slightly fainter magnitude than the 1.5" aperture 
for the F160W filter.}
\end{deluxetable}

\clearpage
%%% TABLE 3
\begin{deluxetable}{lll|ll}
\tablenum{3}
\tablewidth{0pc}
\tablecaption{Mass Limits}
\tablehead{\colhead{Case} & \multicolumn{2}{c|}{z=2.7}  & 
\multicolumn{2}{c}{z=3.9} \\
& \multicolumn{2}{c|}{F300W dropouts} & \multicolumn{2}{c}{F450W dropouts} \\
& \colhead{$M_{\odot}$} & \multicolumn{1}{c|}{$M_{\odot} yr^{-1}$}
& \colhead{$M_{\odot}$} & \colhead{$M_{\odot} yr^{-1}$}}
\startdata 
i & 3.9e8 &  \nodata  &   1.1e9  &  \nodata  \\
ii & 1.5e8 &  \nodata &   1.8e8 &   \nodata  \\
iii & 1.8e7 &  0.36&    6.1e7 &   1.3 
\enddata
\label{t_masslimits}
\end{deluxetable}

\clearpage
%%% TABLE A-1
\begin{deluxetable}{l|lllll}
\tablenum{A-1}
\tablewidth{0pc}
\tablecaption{Expected Observed Wavelengths [\AA]}
\tablehead{& \multicolumn{5}{|c}{Emission Redshift} \\
& \colhead{6.50}  & \colhead{6.25}  & \colhead{6.00}  & \colhead{5.75} & \colhead{5.50}
} 
\startdata
Ly$\alpha$  & 9117  &  8814  & 8510 & 8206  &  7902  \\ 
Ly$\beta$   & 7693  &  7436  & 7180 & 6924  &  6667  \\ 
Ly-limit   &  6838  &  6610  & 6382 & 6154  &  5926   
\enddata
\label{t_a1}
\end{deluxetable}

\clearpage
%%% TABLE A-2
\begin{deluxetable}{l|ll|ll|ll|ll|ll}
\tablenum{A-2}
\tablewidth{0pc}
\tablecaption{Predicted Colors}
\tablehead{& \multicolumn{10}{|c}{Emission Redshift} \\
& \multicolumn{2}{|c}{5.5}  & \multicolumn{2}{c}{5.75}  & \multicolumn{2}{c}{6.0}  
& \multicolumn{2}{c}{6.25}  
& \multicolumn{2}{c}{6.5}  \\
\hline
\multicolumn{1}{c|}{Model}& \colhead{V-I} & \colhead{I-J}  
& \colhead{V-I} & \colhead{I-J}  
& \colhead{V-I} & \colhead{I-J}  
& \colhead{V-I} & \colhead{I-J}  
& \colhead{V-I} & \colhead{I-J}  
} 
\startdata
Uniform      & 1.53&0.47  & 1.63 &0.76 & 1.75&1.09   & 1.75&1.53  & 1.62& 2.12\\ 
Picket Fence & 2.41&0.47  & 3.02 &0.76 & 4.35&1.11   & 5.50&1.63  & 6.81& 2.31\\
No Ly Abs    & 0.71&-0.34 & 0.89&-0.28 & 1.09&-0.23  & 1.29&-0.18 & 1.47&-0.11 
\enddata
\label{t_a2}
\end{deluxetable}

%%%%  FIGURE 1
\begin{figure}
\epsscale{0.85}
\plotone{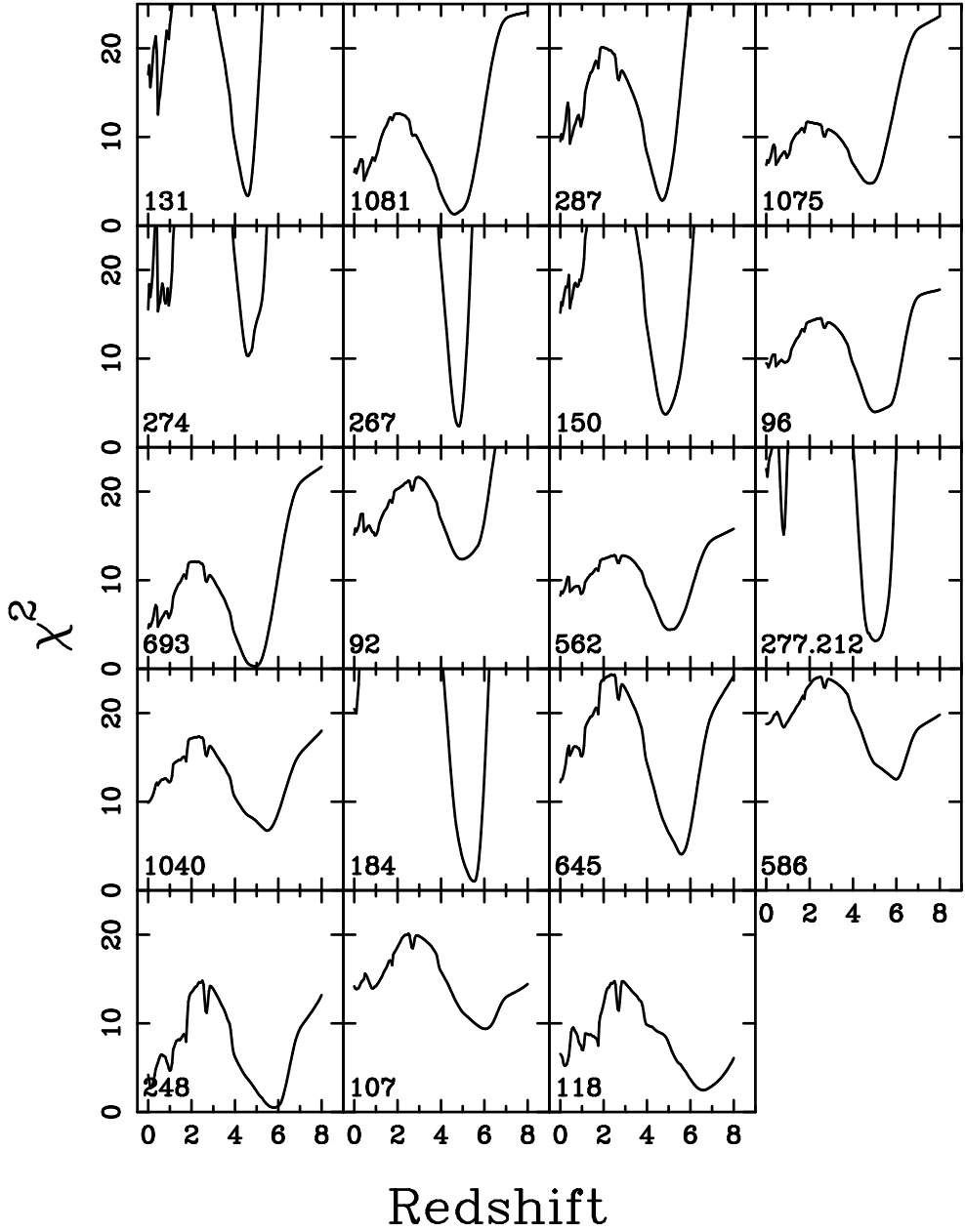}
\caption{The variation of the $\chi^2$ parameter versus redshift for the 19 high redshift candidates. The numbers in each panel refer to the NICMOS identification in the TWS catalog. \label{f_chisq_panel}}
\end{figure}

%%%%  FIGURE 2
%
\begin{figure}
%\epsscale{0.7}
\plotone{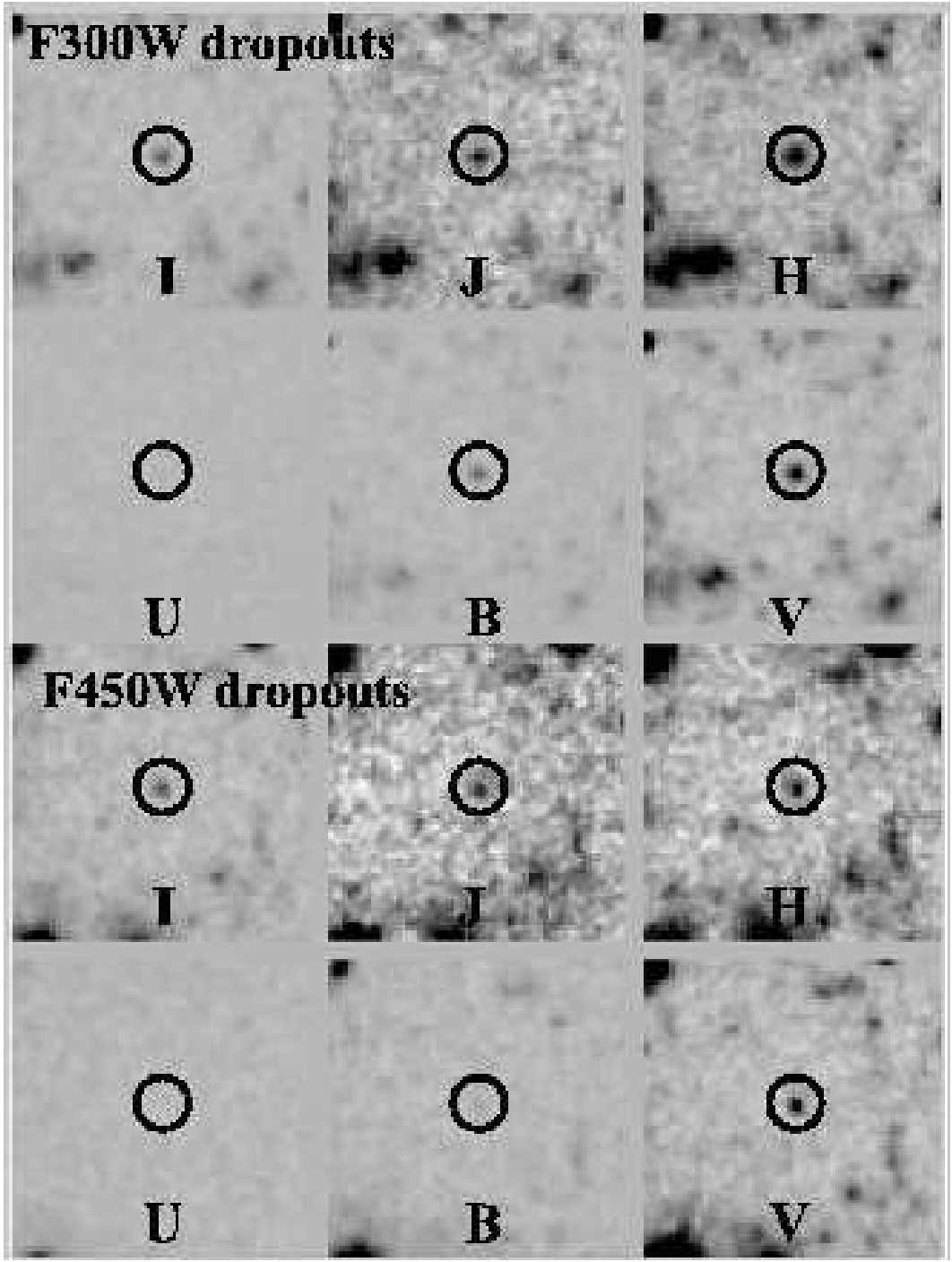}
\caption{
The averaged isolated F300W and F450W dropout galaxies
are shown in the 6 HST wavebands.
The estimated photometric redshifts are $z=2.70$ and
$z=3.9$, respectively. The frames are 6.0 arcseconds on a
side and the averaged galaxies are overlaid with 1.0
arcsecond diameter apertures. 
\label{f_ubmos}}
\end{figure}

%%%%  FIGURE 3 
\begin{figure}
\plotone{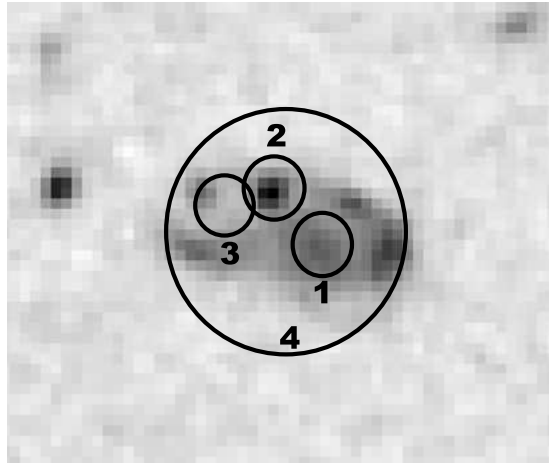}
\caption{The WFPC2 combined F606W+F814W image of WFPC2 4--378.0 whose redshift is 
$z=1.225$ is shown. The frame has been resampled to match the 
orientation and resolution of the NICMOS images. 
The four apertures refer to: 1=bright F160W nucleus,
2=bright F814W off-nuclear knot, 3=dim F814W off-nuclear region,
and 4=the whole galaxy. The flux was measured in 0.6" diameter apertures
for positions 1 through 3 and a 2.6" aperture for position 4.
The photometric redshifts for each of these apertures were $z=1.20$ or $z=1.25$.
\label{f_photz}}
\end{figure}  

%%%%  FIGURE 4 
\begin{figure}
\plotone{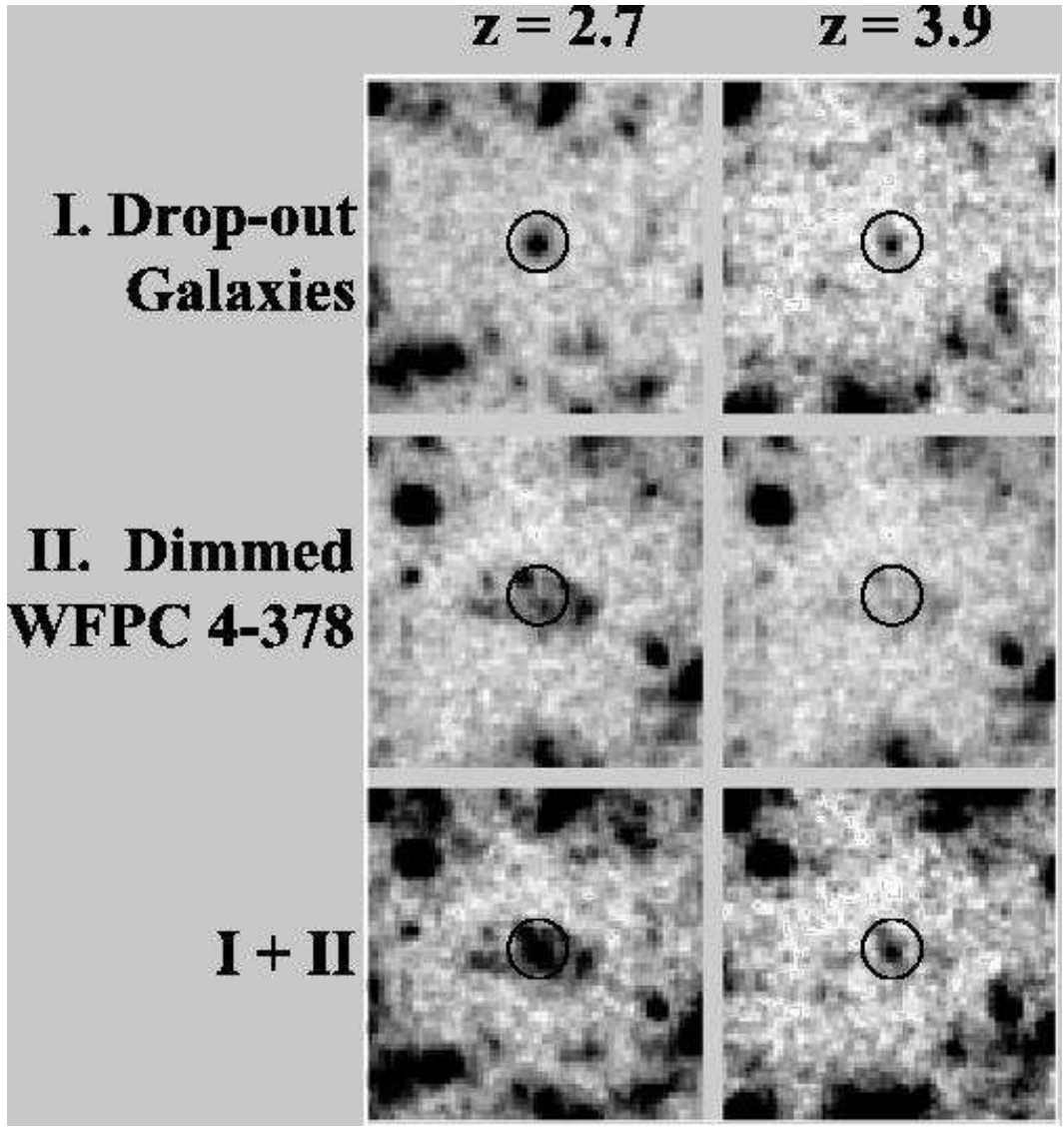}
\caption{The figure simulates what a star--forming galaxy like WFPC2 4--378 
would look like in the F160W 
filter if it were redshifted to $z = 2.7$ (left panels) and $z= 3.9$ 
(right panels). The top
panels (I) reproduce the F300W and F450W dropout galaxies in the F160W band.
The middle panels (II) show the F606W image appropriately
dimmed and scaled as it would approximately appear
in the F160W band at these redshifts, while the bottom panels
(I+II) show the addition of the top and middle panels.
Comparison of the top and bottom panels show that the 
F300W and F450W  drop averaged images lack the extended flux
which an active star--forming galaxy like WFPC2 4--378 shows.
The images are 6 arcseconds on a side and a 1 arcsecond
diameter aperture is overlaid on each image. 
\label{f_compare}}
\end{figure}

%%%  FIGURE 5
\begin{figure}
\plotone{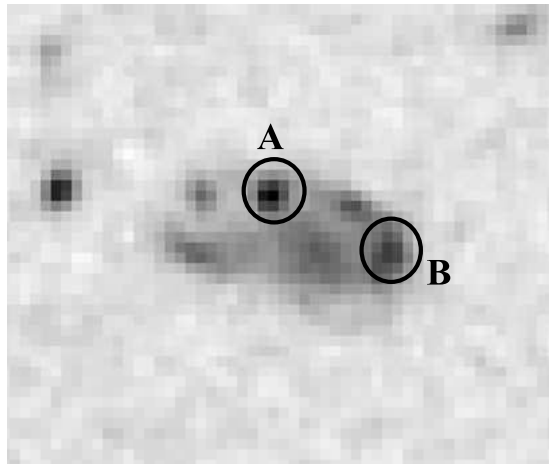}
\caption{This is the same image of WFPC2 4--378.0 as shown in
figure~\ref{f_knots}.
The two bright star-forming knots used as centers of rotation for
the simulations are marked as A and B.
\label{f_knots}}
\end{figure}

%%%%  FIGURE 6 
\begin{figure}
\plotone{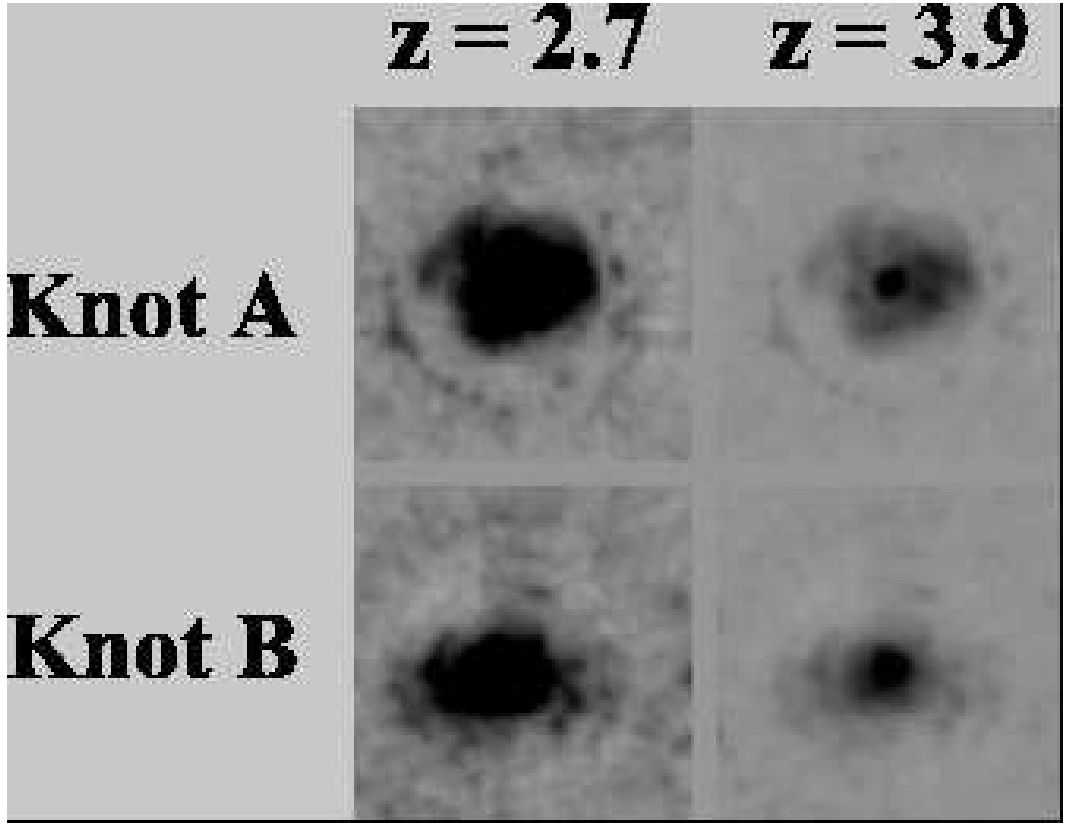}
\caption{The expected appearance of WFPC2 4--378 through the
F160W bandpass at redshifts
of $z=2.7$ (on the left) and $z=3.9$ (on the right) after
summing 20 random rotations about star--forming knots
A (top row) and B (bottom row). 
The extended structure is readily apparent, in contrast to
the F300W and F450W  dropout F160W images in figure~\ref{f_ubmos}.
The images are 6 arcseconds on a side.
\label{f_rotgals}}
\end{figure}

% FIGURE 7
\begin{figure}
\plotone{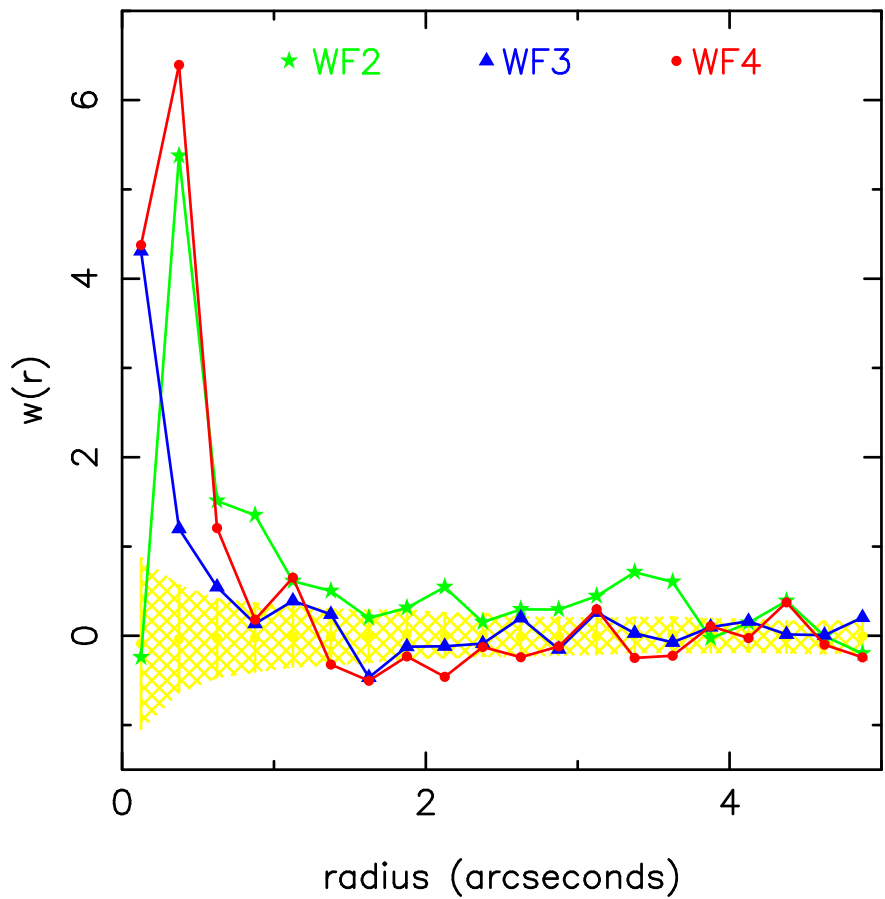}
\caption{The angular correlation function for the 
high redshift color sample of WFPC2 HDF north 
observations in chips 2, 3, and 4, are marked with 
stars, triangles, and circles, respectively. 
The $\pm 1\sigma$ range
for no correlation, determined from
500 realizations of randomly placed points, is shown
as the crosshatched region around $w(r)=0$.
There is a strong signal
for all three WFPC2 chips on scales $r < 0.5"$,
confirming the result found by \cite{Col96}.
The signal is significant at the $3-7\sigma$ level, and
peaks at $0.25" - 0.5"$.   
\label{f_wfpccut}} 
\end{figure}

% FIGURE 8
%
\begin{figure}
\plotone{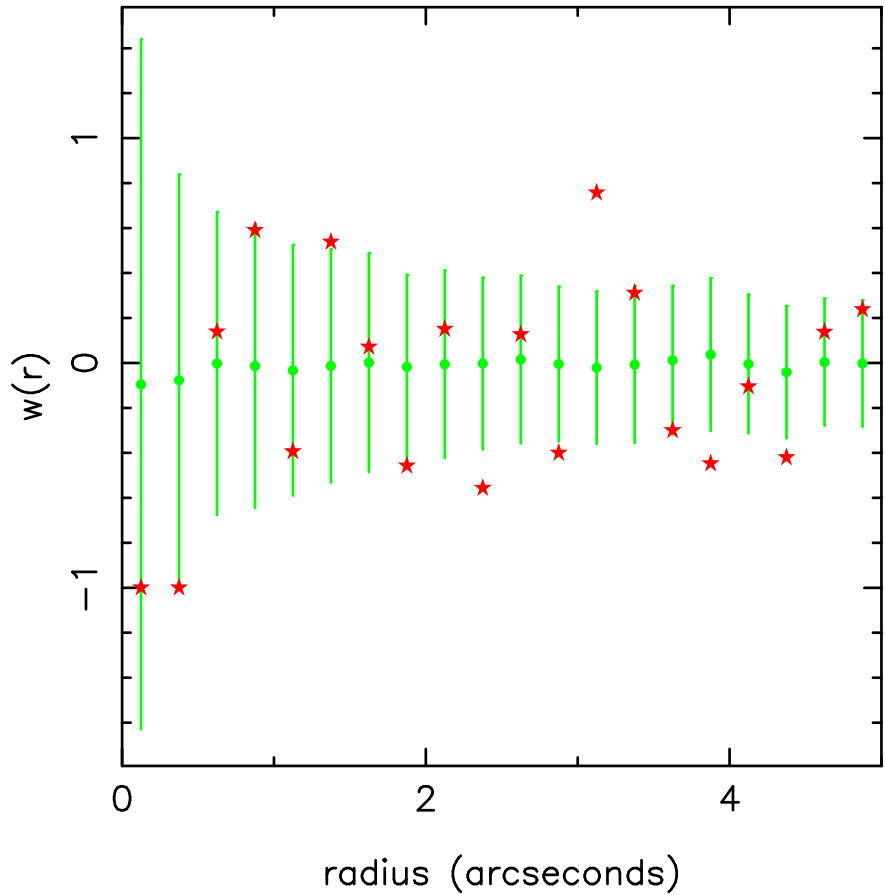}
\caption{The autocorrelation signal for the catalog of the 93 small
galaxies from the NICMOS data set. The data is marked with stars,
the mean in each bin for the random simulations is marked with solid circles, 
and the error bars are $\pm 1 \sigma$ determined from the simulations. 
No correlation sigma is seen, in contrast to the signal
detected in the WFPC2 data set shown in figure~\ref{f_wfpccut}.
\label{f_small}}
\end{figure}

% FIGURE 9
%
\begin{figure}
\plotone{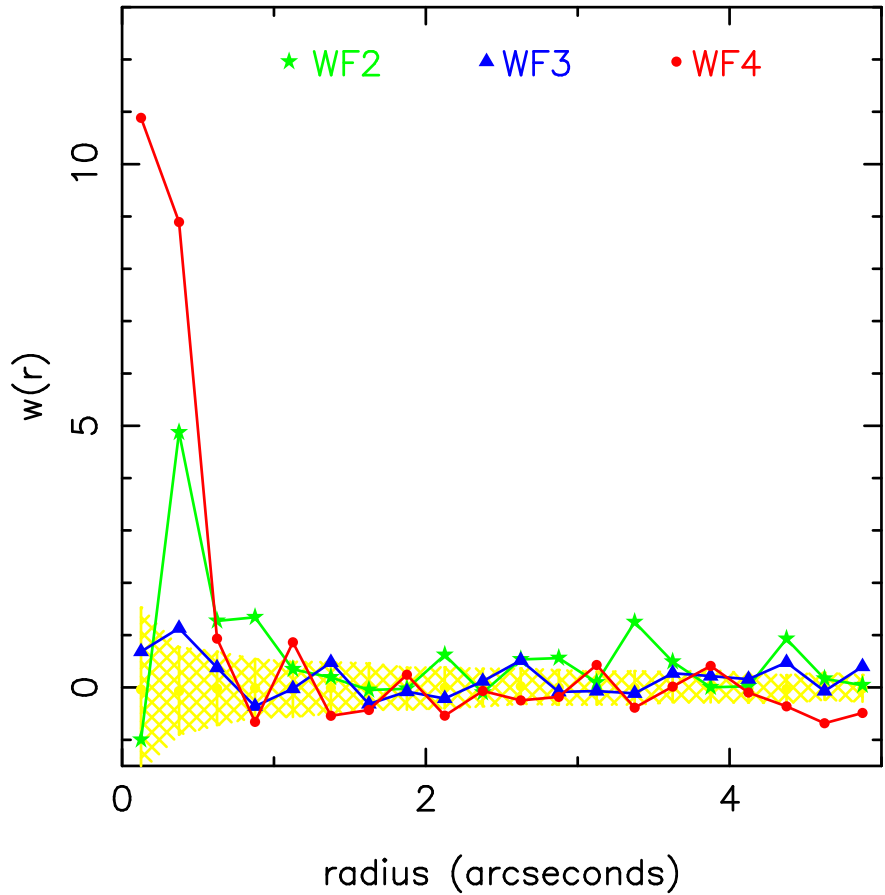}
\caption{The autocorrelation function for 
the same sample as that shown in figure~\ref{f_wfpccut}
but now restricted to  $0.65^{\Box '}$ regions, to match the size of the
NICMOS field. The symbols are as in figure~\ref{f_wfpccut}. The
signal in chip 4 is actually stronger in the smaller field compared
to the full WFPC2 field, about
the same in chip 3, and smaller in chip 2.
\label{f_wfnareacolor}}
\end{figure}

%  FIGURE 10
\begin{figure}
\plotone{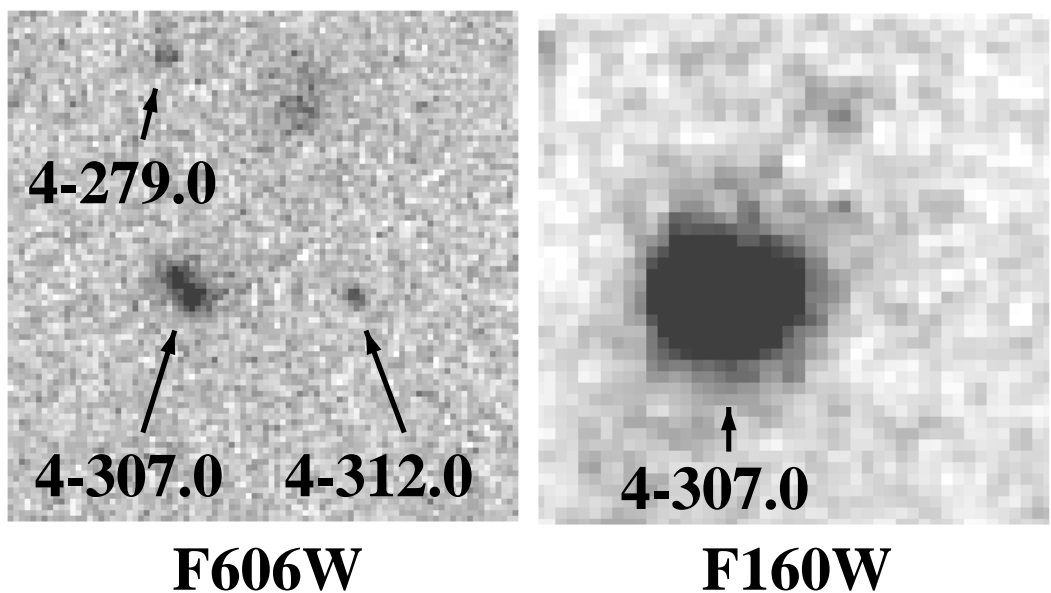}
\caption{The F606W and F160W images of 
a $4.5" \times 4.5"$ region of the northern Hubble Deep Field.
The galaxy just left of the center in both images is
WFPC2 4--307.0 (NIC166.000). WFPC2 4--312.0 is the
galaxy $1.5"$ to the right of 4--307.0 and WFPC2 4--279.0 is
the galaxy
$2.1"$ above 4-307.0 in the F606W image.
We measure 0.6" F606W aperture AB magnitudes of 29.4 and 28.6 
for WFPC2 4--312.0 and 4--279.0, respectively.
These objects are not formally detected by SExtractor in the F160W image
for the S/N thresholds we have adopted, however they have measured
F160W 0.6" aperture magnitudes of 29.7 and 29.6 at the positions 
corresponding to the F606W centroids. 
By contrast, WFPC2 4--307.0 has 0.6" aperture magnitudes in F606W and F160W
of  $= 27.2$ and 23.0. \label{f_bluereddetail}}
\end{figure}

%  FIGURE A-1 (11)
\begin{figure}
\plotone{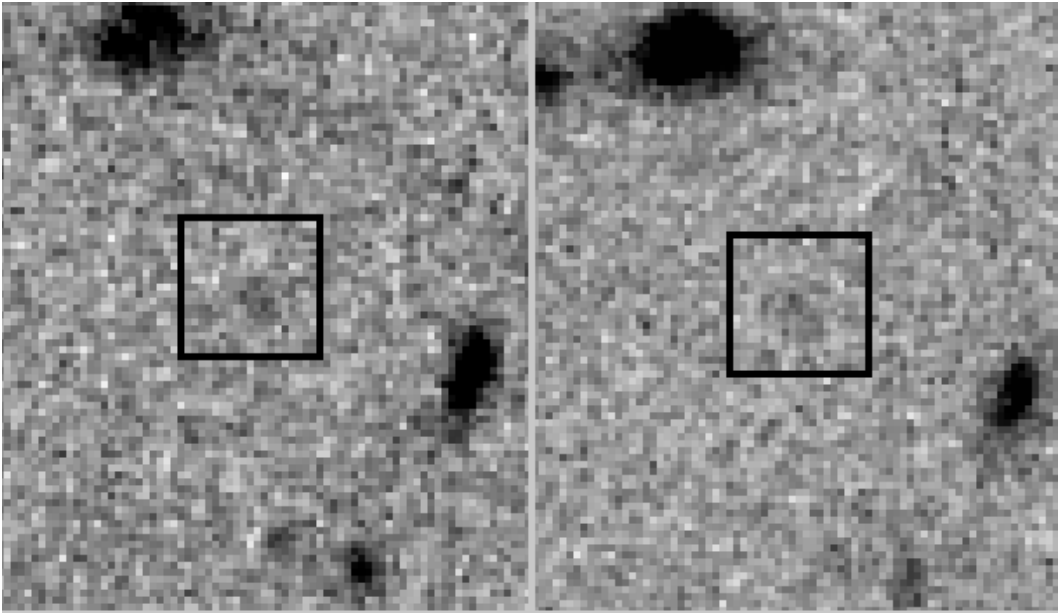}
\caption{ 
The WFPC2 images for HDF 4-601.0 (NIC118.0)
in the F606W (left) and F814W (right) filters are shown. This object is detected 
at a $5.75\sigma$
significance with isophotal AB magnitudes of F814W=30.0 and F606W=30.5 (Williams \etal\ 1996). 
Though faint, there
is clearly flux detected in both wavebands which makes a redshift of $z=6.6$
for this object problematic. There is a 1 arcsecond box centered on the object.
\label{f_append1}}
\end{figure}

%  FIGURE A-2 (12)
\begin{figure}
\plotone{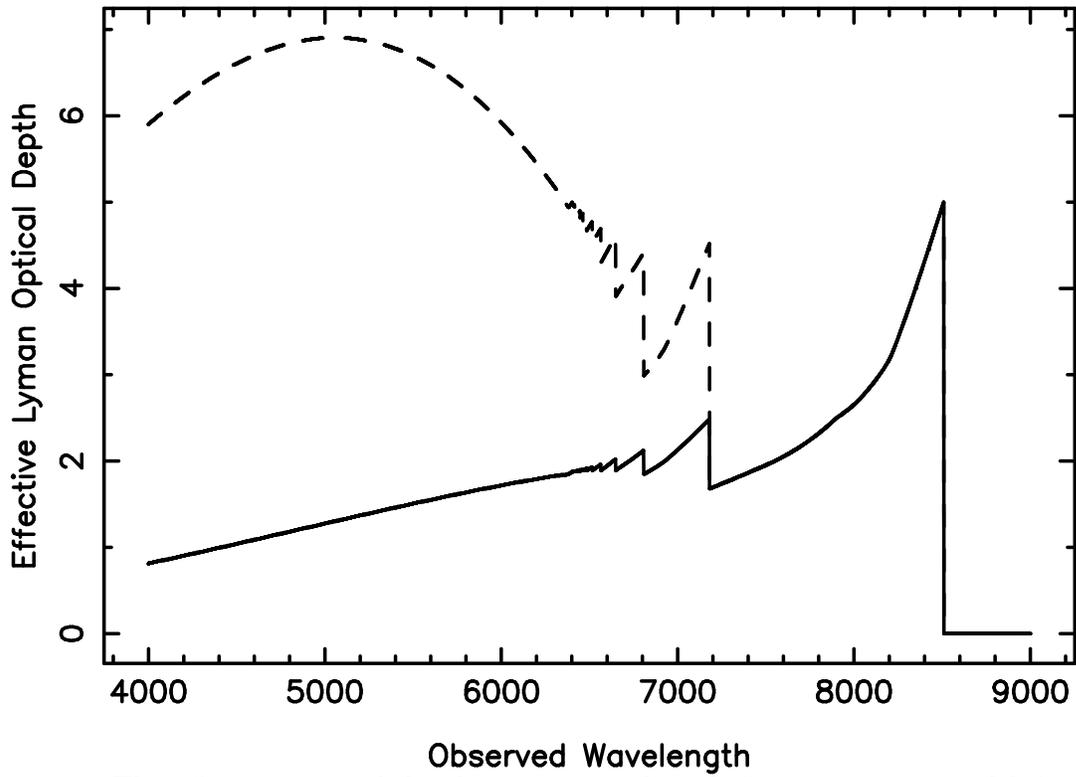}
\caption{ The effective optical depth versus wavelength for an 
emission redshift of z=6 for two models is shown.  The solid line
is for the ``uniform'' model and the dashed line is for the 
``picket fence'' model.  There is no difference in the models
in the region redward of the Ly$\beta$ absorption (7175\AA) but
they vary substantially in the blueward region. 
\label{f_append2}}
\end{figure}


\begin{thebibliography}
\bigskip

\bibitem[Becker \etal\ 2001]{Becker2001} 
Becker, R.H., Fan, X., White, R.L., Strauss, M.A., Narayanan, V.K., 
Lupton, R.H., Gunn, J.E., Annis, J., Bahcall, N.A., Brinkmann, J., Connolly, A.J., 
Csabai, I., Czarapata, P.C., Doi, M., Heckman, T.M., Hennessy, G.S.,
Ivezic, Z., Knapp, G.R., Lamb, D.Q., McKay, T.A., Munn, J.A.,
Nash, T., Nichol, R., Pier, J.R., Richards, G.T., Schneider, D.P., Stoughton, C.,
Szalay, A.S., Thakar, A.R., York, D.G.
2001, \aj,122, 2850

\bibitem[Bertin \& Arnouts\ 1996]{Ber96} Bertin, E., and Arnouts, S. 1996, 
\aap, 117, 393

\bibitem[Colley \etal\ 1996]{Col96} Colley, W.N., Rhoads, J.E., 
Ostriker, J.P., \& Spergel, D.N. 1996, \apj,473, L63 

\bibitem[Cohen \etal\ 2000]{Cohen00} 
Cohen, J.G., Hogg, D.W., Blandford, R., Cowie, L.L., Hu, E., Songaila,
A., Shopbell, P., and Richbert, K. 2000, \apj, 538, 29 

\bibitem[Dickinson 1998]{Dickinson98}
Dickinson, M. 1998, in proceedings of `Space Telescope Science Institute Symposium: 
The Hubble Deep Field,' eds. Livio, M., Fall, S.M., Madau, P.
(Cambridge University Press:New York), 219

\bibitem[Dickinson \etal\ 2000]{Dickinson2000}
Dickinson, M., Hanley, C., Elston, R., Eisenhardt, P.R., 
Stanford, S.A., Adelberger, K.L., Shapley, A., Steidel, C.C., 
Papovich, C., Szalay, A.S., Bershady, M.A., Conselice, C.J., Ferguson,
H.C., Fruchter, A.S. 2000 \apj, 531, 624 

\bibitem[Efstathiou \etal\ 1991]{Efstathiou91} 
Efstathiou, G., Bernstein, G., Katz, N., Tyson, J.A., Guhathakura, P. 
1991, \apj, 380, L47

\bibitem[Fan \etal\ 2001]{Fan2001} 
Fan, X., Narayanan, V.K., Lupton, R.H., Strauss, M.A., 
Knapp, G.R., Becker, R.H., White, R.L., 
Pentericci, L., Leggett, S.K., Haiman, Z.,
Gunn, J.E., Ivezic, Z., Schneider, D.P., Anderson, S.F. 
Brinkmann, J., Bahcall, N.A., Connolly, A.J., 
Csabai, I., Doi, M., Fukugita, M., Geballe, T., Grebel, E.K.,
Harbeck, D., Hennessy, G., Lamb, D.Q., Miknaitis, G., Munn, J.A., 
Nichol, R., Okamura, S., Pier, J.R., Prada, F., Richards, G.T.,
Szalay, A., York, D.G.
2001, \aj,122, 2833

\bibitem[Ferguson 1998]{Ferg98} Ferguson, H.C. 1998,
in proceedings of `Space Telescope Science Institute Symposium: 
The Hubble Deep Field,' eds. Livio, M., Fall, S.M., Madau, P.
(Cambridge University Press:New York), 181

\bibitem[Fernandez-Soto \etal\ 1999]{fly99} 
Fernandez-Soto, A., Lanzetta, K.M., Yahil, A. 1999, 
\apj, 513, 34
                     
\bibitem[Giallongo \etal\ 2002]{Giallongo2002}
Giallongo, E., Cristiani, S., D'Odorico, S., Fontana, A. 2002,
\apj, 568, L9

\bibitem[Haehnelt, Steinmetz \& Rauch 1996]{HSR96}
Haehnelt, M., Steinmetz, M. \& Rauch, M. 1996, \apj, 465, L95

\bibitem[Madau 1995]{madau95} 
Madau, P. 1995,\apj, 441, 18

%\bibitem[Madau, Pozzetti and Dickinson 1998]{madau98} 
%Madau, P., Pozzetti, L.,  Dickinson, M. 1998,\apj, 498, 106

\bibitem[O'Connell \& Marcum 1997]{OM97} O'Connell, R.W.., and Marcum, P. 
1997, in proceedings of `The Hubble Space Telescope and the High 
Redshift Universe,' eds. Tanvir, N.R., Arag\'on-Salamanca, A., \&
Wall, J.V., (World Scientific:London) 63

\bibitem[Rauch, Haehnelt, \& Steinmetz \& Rauch 1997]{RHS97}
Rauch, M., Haehnelt, M. \& Steinmetz, M. 1997, \apj, 481, 601

\bibitem[Roche \etal\ 1996]{Roche96} 
Roche, N., Shanks, T., Metcalfe, N., Fong, R. 1996, 
\mnras, 280, 397

\bibitem[Steidel \etal\ 1996]{Steidel1996}
Steidel, C.C., Giavalisco, M., Dickinson, M., Adelberger, K.L., 1996,
\aj, 112, 352

\bibitem[Steinmetz 1998]{Steinmetz98} Steinmetz, M. 1998,
in proceedings of `Space Telescope Science Institute Symposium: 
The Hubble Deep Field,' eds. Livio, M., Fall, S.M., Madau, P.
(Cambridge University Press:New York), 168


\bibitem[Szalay, Connolly, and Szokoly 1999]{sza99} Szalay, A.J., Connolly,
	A.J., and Szokoly, G.P. 199, \aj, 117, 68


\bibitem[Thompson \etal\ 1998]{niccat} Thompson, R.I., 
Storrie-Lombardi, L.J., Weymann, R.J., Rieke, M.J, Schneider, G., 
Stobie, E. \& Lytle, D. 1998, \aj, 117, 17

\bibitem[Thompson \etal\ 2001]{rit01} Thompson, R.I., 
Weymann, R.J., and Storrie-Lombardi, L.J. 2001, \apj, 546, 694 

\bibitem[Weymann \etal\ 1998]{wey98} Weymann, Ray, J., Stern, D., Bunker, A.,
	Spinrad, H., Chaffee, F.H., Thompson, R.I., and Storrie-Lombardi, L.J.,
	1998, \apj, 505, L95.


\bibitem [Williams \etal\ 1996]{wfcat} Williams, R.E., Blacker, B. Dickinson, M., 
Dixon, W.V.D, Ferguson, H.C., Fruchter, A.S., Giavalisco, M., Gilliland, R.L., 
Heyer, I. Katsanis, R., Levay, Z. Lucas, R.A., McElroy, D.B., Petro, L., Postman, M.,
Adorf, H-M., and Hook, R.N. 1996, \aj, 112, 1335.

\end{thebibliography}
\end{document}